\documentclass[preprint,twocolumn,10pt,a4paper,byrevtex,prl,unsortedaddress,superscriptaddress]{revtex4}

\usepackage{amssymb}
\usepackage{soul}
\usepackage{graphicx}
\usepackage{hyperref}
\usepackage{subfigure}
\usepackage{amsmath}
\usepackage{xcolor}
\usepackage{natbib}
\usepackage{float}
\usepackage{verbatim}
\usepackage{bbold}

\begin{document}

\title{Superconductivity-induced change in magnetic anisotropy in epitaxial ferromagnet-superconductor hybrids with spin-orbit interaction}

\author{C\'esar Gonz\'alez-Ruano}
\affiliation{Departamento F\'isica de la Materia Condensada C-III, Instituto Nicol\'as Cabrera (INC) and  Condensed Matter Physics Institute (IFIMAC), Universidad Aut\'onoma de Madrid, Madrid 28049, Spain}

\author{Lina G. Johnsen}
\affiliation{Center for Quantum Spintronics, Department of Physics, Norwegian University of Science and Technology, NO-7491 Trondheim, Norway}

\author{Diego Caso}
\affiliation{Departamento F\'isica de la Materia Condensada C-III, Instituto Nicol\'as Cabrera (INC) and  Condensed Matter Physics Institute (IFIMAC), Universidad Aut\'onoma de Madrid, Madrid 28049, Spain}

\author{Coriolan Tiusan}
\affiliation{Department of Physics and Chemistry, Center of Superconductivity Spintronics and Surface Science C4S, Technical University of Cluj-Napoca, Cluj-Napoca, 400114, Romania}
\affiliation{Institut Jean Lamour, Nancy Universit\`{e}, 54506 Vandoeuvre-les-Nancy Cedex, France}

\author{Michel Hehn}
\affiliation{Institut Jean Lamour, Nancy Universit\`{e}, 54506 Vandoeuvre-les-Nancy Cedex, France}

\author{Niladri Banerjee}
\affiliation{Department  of  Physics,  Loughborough  University, Epinal  Way, Loughborough, LE11 3TU, United Kingdom}

\author{Jacob Linder}
\affiliation{Center for Quantum Spintronics, Department of Physics, Norwegian University of Science and Technology, NO-7491 Trondheim, Norway}

\author{Farkhad G. Aliev}
\email[e-mail: ]{farkhad.aliev@uam.es}
\affiliation{Departamento F\'isica de la Materia Condensada C-III, Instituto Nicol\'as Cabrera (INC) and  Condensed Matter Physics Institute (IFIMAC), Universidad Aut\'onoma de Madrid, Madrid 28049, Spain}

\begin{abstract}
The interaction between superconductivity and ferromagnetism in thin film superconductor/ferromagnet heterostructures is usually reflected by a change in superconductivity of the S layer set by the magnetic state of the F layers. Here we report the converse effect: transformation of the magnetocrystalline anisotropy of a single Fe(001) layer, and thus its preferred magnetization orientation, driven by the superconductivity of an underlying V layer through a spin-orbit coupled MgO interface.
We attribute this to an additional contribution to the free energy of the ferromagnet arising from the controlled generation of triplet Cooper pairs, which depends on the relative angle between the exchange field of the ferromagnet and the spin-orbit field. This is fundamentally different from the commonly observed magnetic domain modification by Meissner screening or domain wall-vortex interaction and offers the ability to fundamentally tune magnetic anisotropies using superconductivity - a key step in designing future cryogenic magnetic memories.
\end{abstract}

\maketitle

Superconductivity  {(S)} is usually suppressed in the presence of ferromagnetism  {(F)} \cite{Ginsburg1956,Matthias1958,Buzdin1984,Bader2002,Birge2006}. For example, in F/S/F  {spin-valves} the transition temperature $T_\mathrm{C}$ of the S layer is  {different for a} parallel alignment of the F layer moments compared to  {an} anti-parallel alignment  \cite{Tagirov1999,Buzdin1999,Baladie2001,Leksin2011}. Interestingly, for non-collinear alignment of the F layer moments in spin-valves \cite{Leksin2012,Wang2014,Singh2015} or Josephson junctions \cite{Keizer2006,Khaire2010,Robinson2010,Anwar2010,Visani2012,Banerjee2014,Krasnov2014,Linder2015,Eschrig2015,Flokstra2016}, an enhancement in the proximity effect is found due to the generation of long-range triplet Cooper pairs, immune to the pair-breaking exchange field in the F layers.  {So far, the reciprocal modification of the static properties of the ferromagnet by superconductivity has been limited to restructuring \cite{Buzdin1988} and pinning of magnetic domains walls (DWs) by Meissner screening and vortex-mediated pinning of DWs \cite{Bulaevskii2000,Dubonos2002,Fritzsche2009,Curran2015}.} 

 {Modification of the magnetization dynamics in the presence of superconductivity has been studied in \cite{Waintal2002,Tserkovnyak2002,Bell2008,Braude2008,Zhao2008,Konschelle2009,Linder2011,Linder2012,Pugach2012}. Recently, theoretical and experimental results have indicated an underlying role of Rashba spin-orbit coupling (SOC), resulting in an enhancement of the proximity effect and a reduction of the superconducting $T_\mathrm{C}$, along with enhanced spin pumping and Josephson current in systems with a single F layer coupled to Nb through a heavy-metal (Pt) \cite{Bergeret2013,Jacobsen2015,Banerjee2018,Jeon2018,Jeon2019,Satchell2018,Satchell2019}. In this context, V/MgO/Fe \cite{Martinez2020} has been shown to be an effective system to study the effect of SOC in S/F structures with fully epitaxial layers.}

At first glance, altering the  {magnetic order} in S/F heterostructures  {leading to a change in the direction of magnetization} appears non-trivial due to the difference in the energy scales associated with the order parameters. The exchange splitting of the spin-bands and the superconducting gap are about $10^3$ K and $10^1$ K, respectively. However, this fundamentally changes if one considers the possibility of controlling the  {magneto crystalline anisotropy (MCA)} by manipulating the competing anisotropy landscape with superconductivity, since the  {MCA} energy scales are comparable to the superconducting gap energy. Interestingly, emergent triplet superconducting phases in S/SOC/F heterostructures offer the possibility to observe  {MCA modification of} a F layer coupled to a superconductor through a spin-orbit coupled interface, triggered by the superconducting phase \cite{Johnsen2019}.

In this communication, we present evidence that  {cubic in-plane MCA} in V/MgO/Fe(001) system is {modified} by the superconductivity of V through SOC at the MgO/Fe interface \cite{Yang2011}. {Our detailed characterization of the coercive fields of the rotated soft Fe(001) and sensing hard (Fe/Co) ferromagnetic layers by tunnelling magnetoresistance effect (TMR) \cite{Martinez2018} along with numerical simulations dismisses the Meissner screening and DW-vortex interactions as a source of the observed effects.} 

\begin{figure}
\begin{center}
\includegraphics[width=\linewidth]{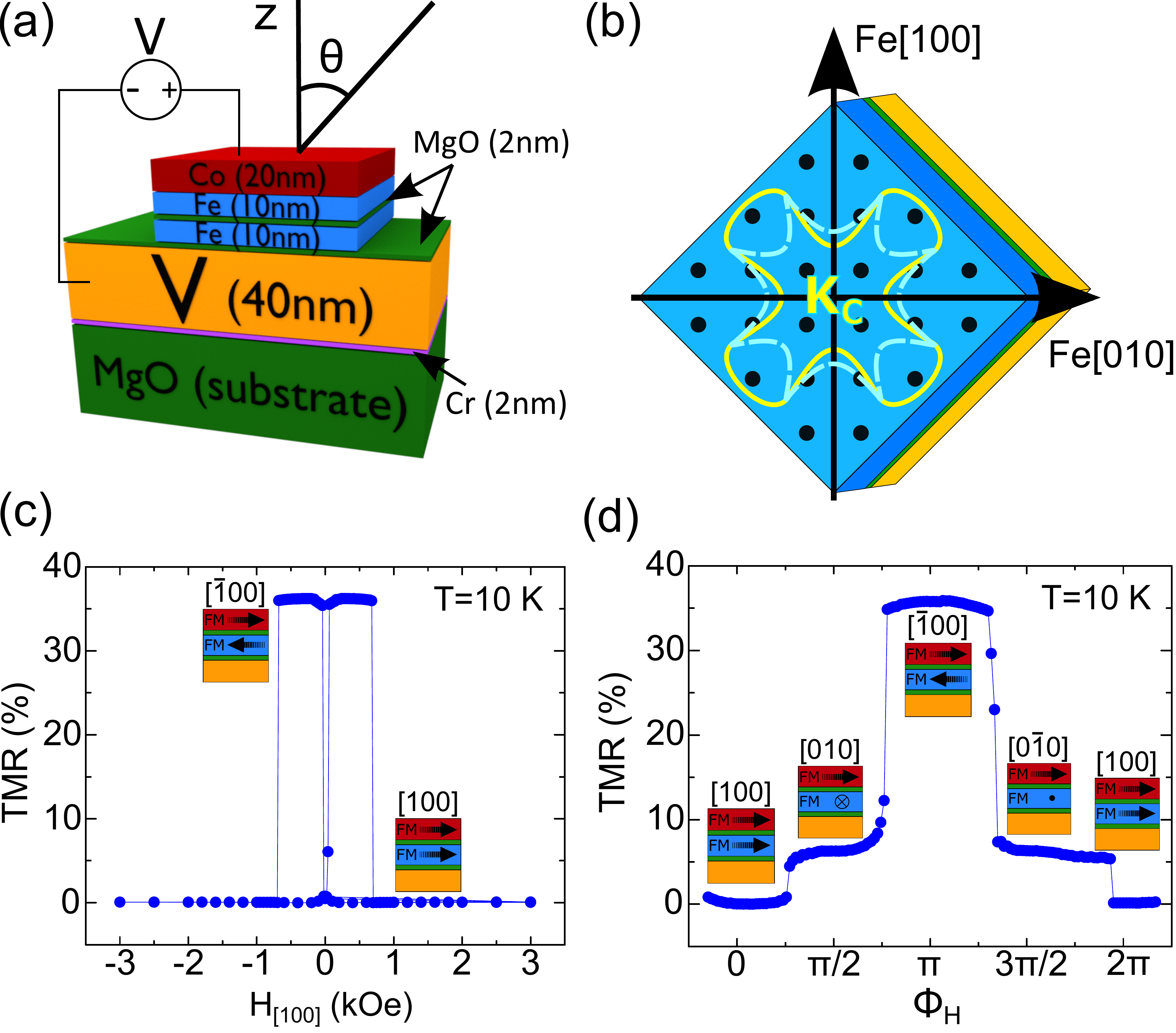}
\caption{\small{(a) Sketch of the junctions under study. Fe(10 nm)Co(20 nm) is the hard (sensing) layer while Fe(10 nm) is the soft ferromagnet where spin reorientation transitions are investigated. (b), Sketch showing the top view without the hard Fe/Co layer, with the 4-fold in plane magnetic energy anisotropy expected for the Fe(001) atomic plane of the magnetically free layer, for temperatures above $T_\mathrm{C}$ (yellow line) and well below $T_\mathrm{C}$ (dashed cyan). Note that during the epitaxial growth, the Fe lattice is rotated by 45 degress with respect to MgO. Parts (c) and (d) show in-plane spin reorientation transitions between parallel (P), perpendicular in plane (PIP) and antiparallel (AP) relative magnetization allignments of the soft and hard F layers for a $30\times30$ $\mu\text{m}^2$ junction at T=10 K (above $T_\mathrm{C}$). Indices above the inset sketches indicate the direction of the soft layer. The in-plane rotation has been carried out with the angle $\Phi_H$ of the magnetic field relative to the Fe[100] axis going from $-30$ to $390$ degrees.}}
\label{Fig1}
\end{center}
\end{figure}

The  {magnetic tunnel junction (MTJ)} multilayer stacks have been grown by molecular beam epitaxy (MBE) in a chamber with a base pressure of $5\times10^{-11}$ mbar following the procedure described in \cite{Tiusan2007}. The samples were grown on [001] MgO substrates. Then a 10 nm thick seed of anti-diffusion MgO underlayer is grown on the substrate to trap the C from it before the deposition of the Fe (or V). Then the MgO insulating layer is epitaxially grown by e-beam evaporation, the thickness approximately $\sim 2$ nm and so on with the rest of the layers. Each layer is annealed at 450 ºC for 20 mins for flattening. After the MBE growth, all the MTJ multilayer stacks are patterned in 10-40 micrometre-sized square junctions (with diagonal along [100]) by UV lithography and Ar ion etching, controlled step-by-step \textit{in situ} by Auger spectroscopy. 
The measurements are performed inside a JANIS$^{\tiny{\textregistered}}$ He$^3$ cryostat. The magnetic field is varied using a 3D vector magnet. 
 {For the in-plane rotations, the magnetic field magnitude was kept at 70-120 Oe, far away from the soft Fe(001) and hard Fe/Co layers switching fields obtained from in-plane TMRs (see Supplemental Material S1,S2 \cite{SuppMat}). This way, only the soft layer is rotated and the difference in resistance can be atributted to the angle between the soft and hard layers.}

Figure \ref{Fig1}a shows the device structure with the Fe/Co hard layer sensing the magnetization alignment of the 10-nm thick Fe(001) soft layer. A typical TMR plot above $T_\mathrm{C}$ is shown in Figure \ref{Fig1}c. The resistance switching shows a standard TMR between the P and AP states. However, the epitaxial Fe(001) has a four-fold in-plane anisotropy with two ortogonal easy axes - [100] and [010] - (Figure \ref{Fig1}b). These MCA states could be accessed by an in-plane rotation of the Fe(001) layer with respect to the Fe/Co layer using field greater than the coercive field of the Fe(001) layer without disrupting the Fe/Co magnetization  {(see also Supplemental material S1 \cite{SuppMat} for the magnetic characterization of the Fe/Co layer)}. This is shown in Figure \ref{Fig1}d, where TMR is plotted as a function of the in-plane field angle with respect to the [100] direction angle $\Phi_H$. This gives rise to four distinct magnetization states with P, perpendicular in-plane (PIP) and AP states reflected by the TMR values.   {Supplemental Material S3 \cite{SuppMat} discusses the weak magnetostatic coupling between the two FM layers (detected through resistances in-between the P and AP states in the virgin state of different samples), showing that it does not affect the capability to reorient the soft layer independently of the hard one. It also demonstrates that the soft layer retains different magnetic directions at zero field.}

\begin{figure}
\begin{center}
\includegraphics[width=\linewidth]{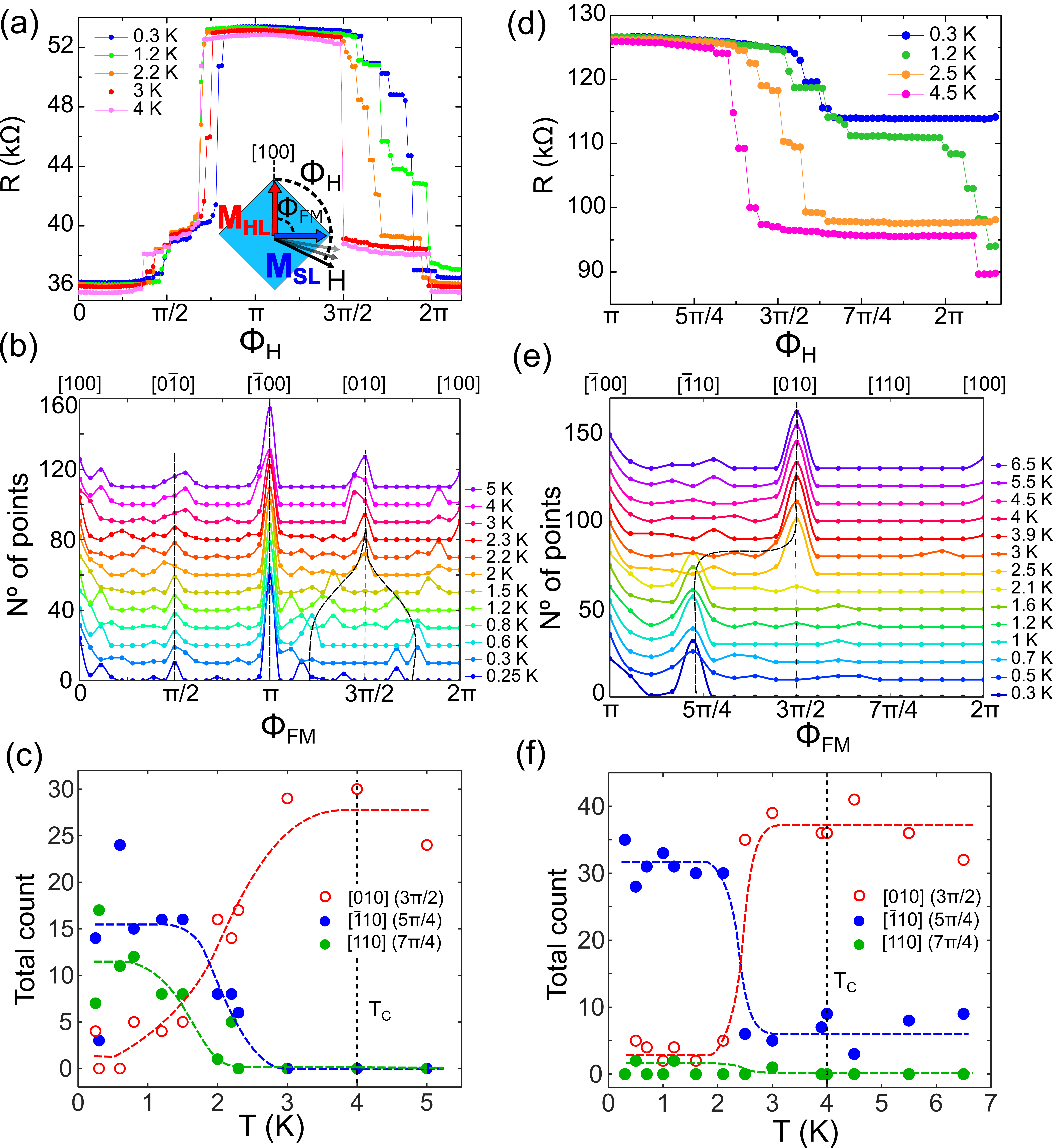}
\caption{\small{ {Typical angular dependence of the resistance of a V/MgO/Fe/MgO/Fe/Co junction on the orientation of the in plane field with respect to the main crystalline axes from above to below $T_\mathrm{C}$ when the rotation is initialted from P (a-c) and from AP state (d-f). The inset sketches the experimental configuration, showing the angles between the ferromagnetic layers ($\Phi_{FM}$) and of the external magnetic field ($\Phi_H$). 
Parts (b,e) correspondingly represent the experimental data shown in (a,d) in form of histograms, dividing the 0-$2\pi$ interval in 36 zones. Parts (c,f) plot the histograms in (b,e) as counts vs temperature for the intermediate states (AP$+\pi/4$ or the $[\overline{1}10]$ axis, AP$+\pi/2=$PIP or [010], and AP$+3\pi/4$ or [110]) for the second half of the rotation.}}}
\label{Fig2}
\end{center}
\end{figure}

Figure \ref{Fig2}  {analyzes the most probable  {in-plane} magnetization orientations of the Fe(001) layer through magnetic field rotations} at fixed temperatures from above to below $T_\mathrm{C}$.  {Typically, no qualitative changes in TMR are observed above and below $T_{\mathrm{C}}$ in the $0-\pi$ field rotation angle ($\Phi_H$) span (Figure \ref{Fig2}a)}. However, in the $\pi-2\pi$ range,  {the TMR qualitatively changes below $T_\mathrm{C}/2$, possibly indicating new stable magnetization states along 
different directions to the ones stablished by the principal crystallographic axes (Figure \ref{Fig2}a).}

To ascertain the exact angle $\Phi_{FM}$ between the two F layers, we have calibrated the magnetization direction of the soft layer with respect to the hard Fe/Co using the Slonczewski formula (Supplemental Material S4 \cite{SuppMat}).  {The aplicability of the macrospin approach to describe TMRs and magnetization reorientation resides in the high effective spin polarization obtained ($P=0.7$) \cite{Martinez2018}, approaching to the values typically reported for Fe/MgO in a fully saturated state \cite{Parkin2004,Yuasa2004}.} Figure \ref{Fig2}b is a histogram representing the probability of obtaining a specific $\Phi_{FM}$ as temperature is lowered from above to below $T_\mathrm{C}$. We observe that the most probable Fe(001) directions are oriented along the [100] and [010] principal axes above $T_\mathrm{C}/2$, while below $T_\mathrm{C}/2$ it splits in  {three} branches roughly oriented along $\pi/4$ angles.  {The split of the $[0\overline{1}0]$ state into three branches is also visualized in Fig.\ref{Fig2}e, with a plot of the counts vs. temperature around the $[\overline{1}\overline{1}0], [0\overline{1}0]$ and $[1\overline{1}0]$ magnetization directions.}

Interestingly, once the rotation is initiated in the AP configuration, the magnetization apparently locks in the ($\pi+\pi/4$)  {(or $[\overline{1}\overline{1}0]$)} state (Figures \ref{Fig2}b,d,f). This probably arises due to the improved initial macrospin alignment, which is not  {fully} achieved in the  {AP state with a preceeding} P-AP rotation.  {We believe that with the full 2$\pi$ field rotation, magnetization inhomogeneities or local DWs created during the P-AP state rotation help to overcome MCA energy barriers more easily. The suggested suppresion of the local DWs with the magnetization rotation initated from the AP state can be indirectly inferred from the broadening of the $[\overline{1}00]$ to $[0\overline{1}0]$ transition in the normal state detected as a small (extrinsic) number of counts around $[\overline{1}10]$ (Figure \ref{Fig2}f).}


 {For a more systematic analysis}, we performed a series of in-plane TMR measurements along different directions relative to the symmetry axes.  {The first experiment (i) was performed with an initial saturation field of $\pm1$ kOe in the [100] direction, followed by a TMR in the [210] direction (between [100] and [110]). The second (ii) initially saturates both the hard and soft layers along the [100] direction. Then, a minor loop is performed starting from zero field and going up to 150 Oe along the [110] axis}.

\begin{figure}
\begin{center}
\includegraphics[width=\linewidth]{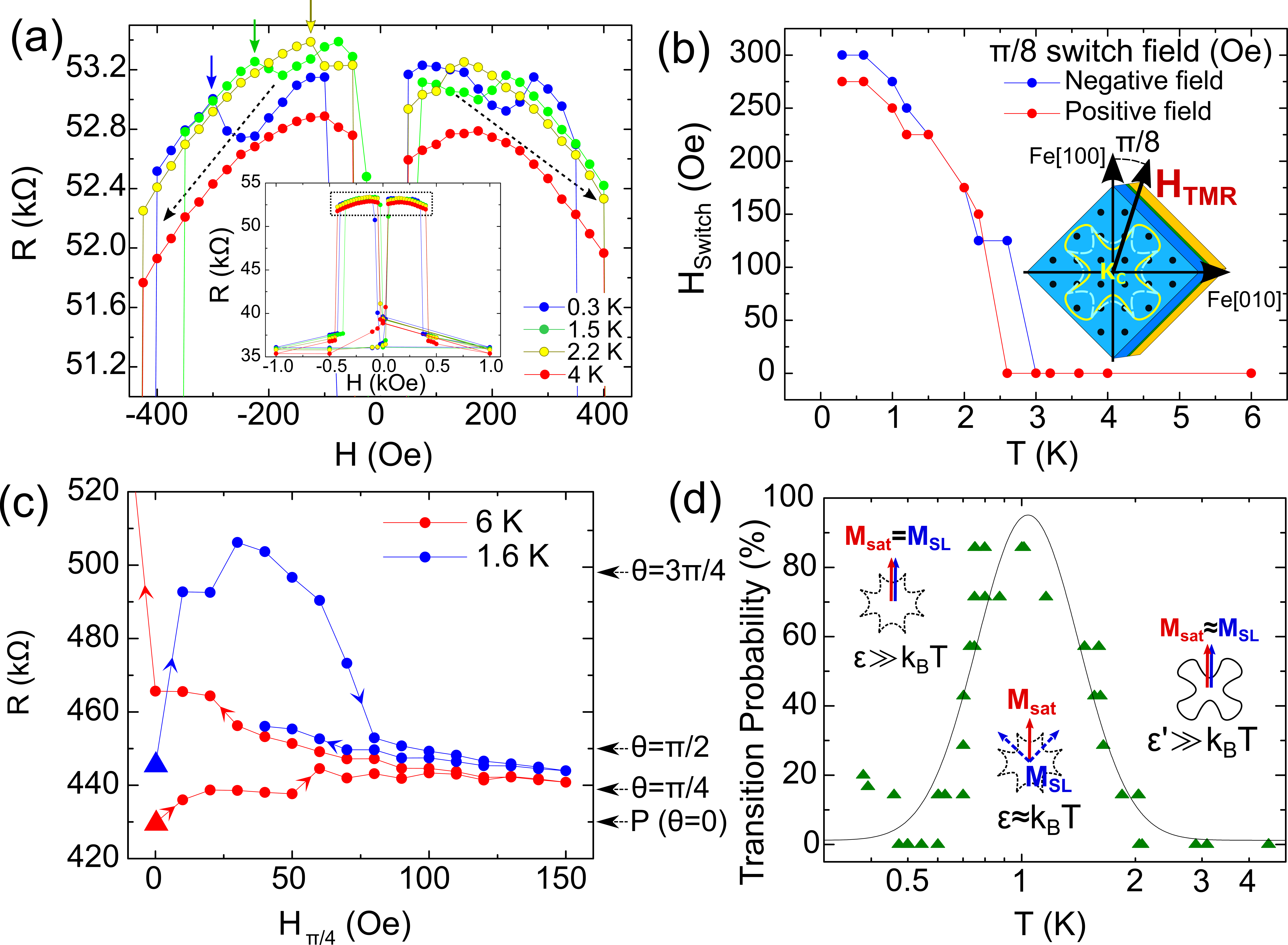}
\caption{\small{(a) TMR measurements on a S/F/F $30\times30$ $\mu\text{m}^2$ junction with H oriented along  {[210]} (inset in (b)), for various temperatures. The increase in R is associated with a transition from the   {[110]} magnetization orientation to a forced  {[210]} direction of the soft layer. (b), Variation of the transition field with T for the positive and negative field branches. Inset: exchange energy anisotropy and direction of the applied field (H$_{\text{TMR}}$). (c), Two TMRs performed on a $10\times10$  $\mu\text{m}^2$ junction in the [110] direction at T=6 K and 1.6 K, after applying 1 kOe in the [100] direction. The 6 K TMR starts in P state, while the 1.6 K TMR starts already in a tilted state. Right axis: estimation of the angle $\theta$ between the two F layers based on the Slonczewski formula. (d), Probability of finding a tilted state at $H=0$ (triangular points in (b)) vs T (in log scale), averaged with 7 experimental points around each T. The line is a guide for the eye. Insets: sketch of the magnetic anisotropy below and above $T_\mathrm{C}$, with the saturation magnetization (M$_{\text{sat}}$) and the zero field magnetization state measured for the soft layer (M$_{\text{SL}}$). $\varepsilon$ and $\varepsilon'$ represent the energy barrier separating the [100] magnetization direction from the closest minimum below and above $T_\mathrm{C}$, respectively.}}
\label{Fig3}
\end{center}
\end{figure}

{Both experiments further suggest the possibility of superconductivity-induced changes of MCA.} The inset of Figure \ref{Fig3}a shows the full field sweep range in the  {first (i)} configuration, and Figure \ref{Fig3}a zooms in close to the AP configuration. When we sweep the field  {in the [210] direction}, we  {detect a weak but robust resistance upturn} at temperatures below approximately $T_\mathrm{C}/2$ (Figure \ref{Fig3}). This additional TMR increase (shown by the arrows in Figure \ref{Fig3}a) roughly corresponds to an 8-10 degree rotation in the relative spin direction between the soft and hard layer towards their AP alignment  {(see Supplemental Material S4 \cite{SuppMat} for an analysys of the calculated angle error)}.  {Within the proposed macrospin approximation,} this could be  {understood} as a redirection of the soft layer magnetization forced by the external field, from the initially blocked [110] direction towards the external field  {[210]} direction. A strong increase of the characteristic field, $H_{switch}$, required to reorient the soft layer from [110] towards  {[210]} when T decreases below $T_\mathrm{C}/2$,  {could} reflect the superconductivity-induced  {MCA} energy minimum along the [110] direction.

The minor TMR loops along [110] (Figure \ref{Fig3}c)  {realized after saturation along [100]} point on a thermally induced magnetization reorientation  {from [100]} towards [110] even at zero field, in a temperature range below $T_\mathrm{C}$ where the barrier between adjacent energy minima is comparable to $k_BT$. The zero-field reorientation becomes less probable when the thermal energy is insufficient to overcome the barrier (Figure \ref{Fig3}d). An estimation of the in-plane normal-state  {MCA energy barrier} done through magnetization saturation along [100] and [110] provides a value of only a few $\mu\text{eV}$/atom (Supplemental Material S5 \cite{SuppMat}). However, the real barrier is determined by the nucleation volume, which depends on the exchange length in the material. With a DW width of about 3 nm for Fe(001) we estimate the  {MCA} barrier to be at least  $10^0-10^1$ mV.


Before  {describing our explanation of the MCA modification of Fe(001) in the superconducting state of V(40 nm)/MgO(2 nm)/Fe(10 nm) system, we discard alternative interpretations of the observed effects. Meissner screening \cite{Bulaevskii2000,Dubonos2002}, if present, would introduce about a 10$\%$ correction to the actual magnetic field independently of the external field direction (see Supplemental Material S2 \cite{SuppMat}). The reason for the weak in-plane field screening could be the small superconductor thickness (40 nm), only slightly exceeding the estimated coherence length (26 nm). On the other hand, intermediate multidomain states are expected to be absent in when magnetization is directed along [110] (Supplemental material S6 \cite{SuppMat}).  {Indeed, our experiments show that magnetization, when locked below $T_\mathrm{C}$ in the ($\pi+\pi/4$) angle, hardly depends on the absolute value of the external field along [110] varied between 0 and 100 Oe.} Moreover, simulations of the vortex-DW interaction using MuMax3 \cite{MuMax3} and TDGL codes \cite{Lara2020} discard the vortex mediated DW pinning \cite{Fritzsche2009,Curran2015} scenario  {including when interfacial magnetic defects created by misfit dislocations \cite{Herranz2010} are considered (see Supplemental material S6 \cite{SuppMat})}. The vortex pinning mechanism also contradicts that only the $(0-\pi)$ field rotation span (Figure \ref{Fig2}a) gets affected below $T_\mathrm{C}/2$. The observed irrelevance of the junction area (Supplemental material S7 \cite{SuppMat}) contradics the importance of the vortex-edge DWs interaction. The  shape and vortex-DWs interaction effects, if relevant, would strengthen magnetization pinning along [100], but not [110] (Supplemental material S6, S7 \cite{SuppMat})}. Finally, we also indicate that the MCA modification from singlet superconductivity would not enable any  {zero field} rotation to non-collinear misalignment angles, in contrast to our data (Fig. \ref{Fig3}d).


\begin{figure}
\begin{center}
\includegraphics[width=\linewidth]{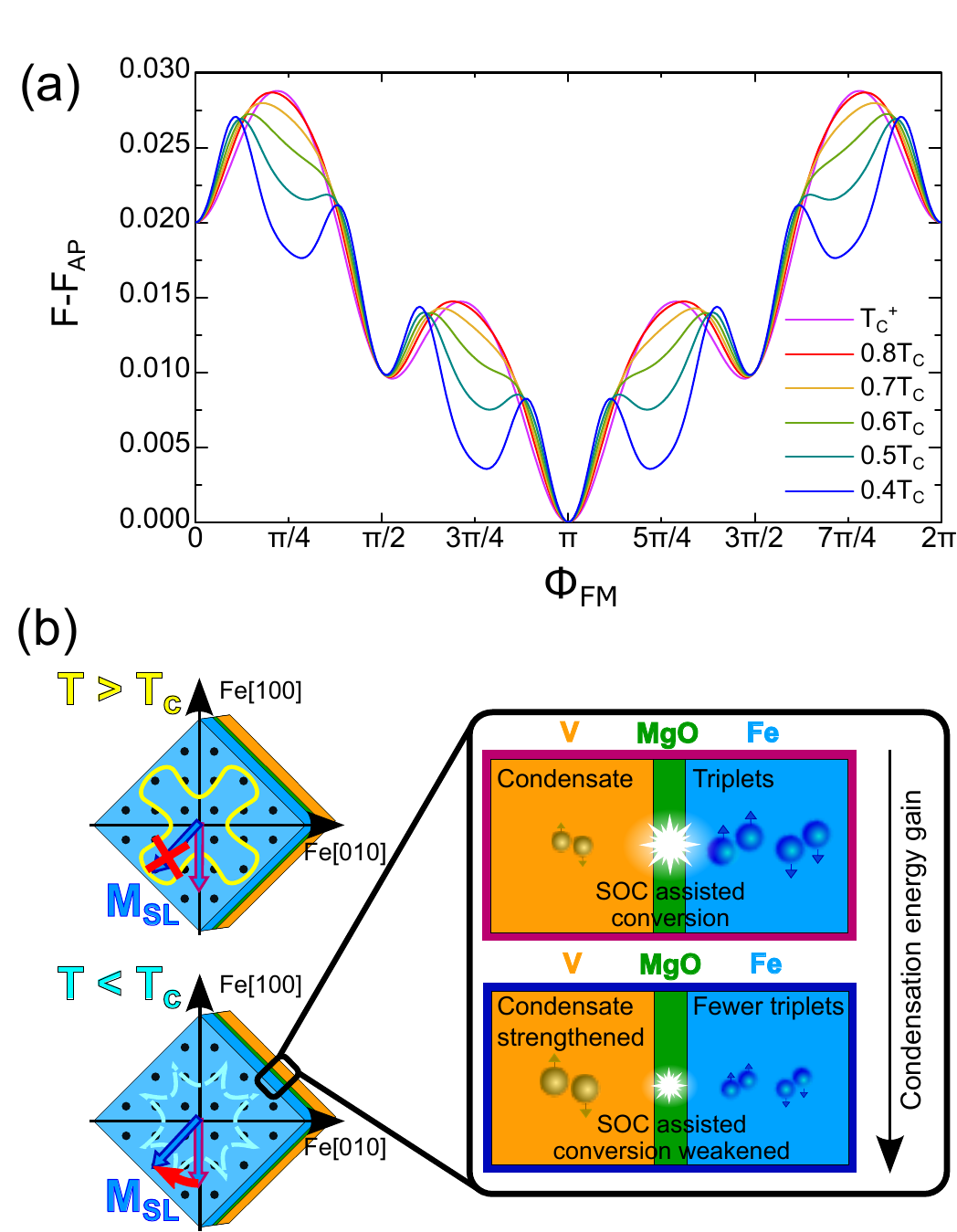}
\caption{\small{Numerical modelling. (a), Free energy $F$ vs in-plane magnetization angle $\Phi_{{FM}}$ for temperatures below the superconducting critical temperature and just above the critical temperature ($T_\mathrm{C}^+$). The free energy is plotted relative to the free energy in the AP configuration $F_{AP}$ and has been normalized to the hopping parameter $t$ used in the tight-binding model. (b). Illustration of the physical origin of the change in magnetic anisotropy induced by the superconducting layer. Above $T_\mathrm{C}$, V is a normal metal and the soft Fe layer has a 4-fold in-plane magnetic energy anisotropy (yellow line). Below $T_\mathrm{C}$, V is superconducting and influences the soft Fe layer via the proximity effect: a leakage of Cooper pairs into the ferromagnet. Due to the  SOC at the interface, a magnetization-orientation dependent generation of triplet Cooper pairs occurs. The generation of triplets is at its weakest for a magnetization pointing in the $[\overline{1}\overline{1}0]$ direction, giving a maximum for the superconducting condensation energy gain. This modifies the magnetic anisotropy of the soft Fe layer (cyan line), enabling magnetization switching to the $[\overline{1}\overline{1}0]$ direction (blue arrow). The magnetic anisotropy does not show the  {weak} AP coupling between the two Fe layers, causing an absolute minimum in $\Phi_{FM} = \pi$ (a).}}
\label{Fig4}
\end{center}
\end{figure}

 {To explain our results, we consider the possibility in which the invariance of the superconducting proximity effect to magnetization rotation is broken in the presence of SOC. It has been predicted that triplet-superconductivity is effectively generated even for weakly spin-polarized ferromagnets with a small spin-orbit field \cite{Vezin2020}. In addition to generating triplet pairs, the SOC}  {also introduces an angle-dependent anisotropic depairing field for the triplets \cite{Jacobsen2015,Johnsen2019}.}  {In V/MgO/Fe, the Rashba field is caused by a structural broken inversion symmetry at the MgO interfaces \cite{Martinez2020}}.
 {We model our experimental results using a tight-binding Bogolioubov-de Gennes Hamiltonian on a lattice and compute the free energy (Supplemental material S8 \cite{SuppMat}). The Hamiltonian includes electron hopping in and between the different layers, a Rashba-like  SOC at the MgO/Fe interface, an exchange splitting between spins in the Fe layers, and conventional $s$-wave superconductivity in the V layer.}
 {The free energy determined from this Hamiltonian includes the contribution from the superconducting proximity effect, and an effective in-plane magnetocrystalline anisotropy favoring magnetization along the [100] and [010] axes.}
Experimentally, we see a weak anti-ferromagnetic coupling between the Fe(100) and Fe/Co layers  {(which does not affect the capability to reorient the soft layer independently of the hard one)} described by an additional contribution $f_{{AF}}\cos(\Phi_{FM})$ with a constant parameter $f_{{AF}}>0$. 

Figure \ref{Fig4} shows the total free energy 
of the system as a function of the IP magnetization angle $\Phi_{FM}$ for decreasing temperatures.  
 {Due to the increase in the superconducting proximity effect, additional local minima appear at $\Phi_{FM} = n\pi/2+\pi/4$, where $n=0,1,2,...$ (\textit{i.e.} $[110]$, $[\Bar{1}10]$, $[\Bar{1}\bar{1}0]$, and $[1\Bar{1}0]$, respectively).}
This is a clear signature for the proximity-induced triplet correlations.  {These are most efficiently generated at angles $\Phi_{FM} = n\pi/2$ (\textit{i.e.} $[100]$, $[010]$, $[\Bar{1}00]$, and $[0\Bar{1}0]$) for a heterostructure with a magnetic layer that has a cubic crystal structure like Fe \cite{Johnsen2019}.} As a result, the decrease in the free energy is stronger at angles $\Phi_{FM} = n\pi/2+\pi/4$ where more singlet Cooper pairs survive.  
Our numerical results thus confirm that the experimentally observed modification of the anisotropy can be explained by the presence of SOC in the S/F structure alone, without including superconducting proximity effects from misalignment between the Fe(100) and Fe/Co layers.
Moreover, Figure \ref{Fig4} illustrates why the $\Phi_{FM} = n\pi/2+\pi/4$ states only appear experimentally when the external field is rotated from an AP to P alignment (Figure \ref{Fig2}). Because of the  {weak} AP coupling between the ferromagnetic layers, the energy thresholds for reorienting the magnetization from one local minimum to the next are higher under a rotation from AP to P alignment.

In conclusion,  {we present experimental evidence for superconductivity-induced change in magnetic anisotropy in epitaxial ferromagnet-superconductor hybrids with spin-orbit interaction.}   {We believe that this mechanism is fundamentally different from the previous reports of magnetisation modification arising from Meissner screening and vortex induced domain wall pinning, even though the spin-triplet mechanism and performed simulations require many assumptions.} Our results establish superconductors as tunable sources of magnetic anisotropies and active ingredients for future low dissipation  {superspintronic} technologies.   {Specifically, they could provide an opportunity to employ spin-orbit proximity effects in magnetic Josephson junction technology and appoach it to Fe/MgO-based junctions that are widely used in commercial spintronic applications.}

\section*{Acknowledgements}
We acknowledge Mairbek Chshiev and Antonio Lara for help with simulations, Yuan Lu for help in sample preparations and Igor Zutic and Alexandre Buzdin for the discussions. The work in Madrid was supported by Spanish Ministerio de Ciencia (MAT2015-66000-P, RTI2018-095303-B-C55, EUIN2017-87474) and Consejer\'ia de Educaci\'on e Investigaci\'on de la Comunidad de Madrid (NANOMAGCOST-CM Ref. P2018/NMT-4321) Grants. FGA acknowledges financial support from the Spanish Ministry of Science and Innovation, through the “María de Maeztu” Programme for Units of Excellence in $R\&D$ (MDM-2014-0377, CEX2018-000805-M). NB was supported by EPSRC through the New Investigator Grant EP/S016430/1. The work in Norway was supported by the Research Council of Norway through its Centres of Excellence funding scheme grant 262633 QuSpin. C.T. acknowledges ``EMERSPIN'' grant ID PN-IIIP4-ID-PCE-2016-0143, No. UEFISCDI: 22/12.07.2017. The work in Nancy was supported by CPER MatDS and the French PIA project ``Lorraine Universit\'e d'Excellence'', reference ANR-15-IDEX-04-LUE.

\section{Supplementary Material}

In the supplementary material, {the section S1 presents a magnetic characterization of the hard Fe/Co layer of the junctions under study. Section S2 presents a magnetic characterization of the soft Fe(001) layer and studies the possible influence of the Meissner screening on the coercive fields of the soft and hard layers. Section S3 estimates the strength of the weak antiferromagnetic coupling between magnetically soft and hard electrodes.} Section S4 provides details about the calibration of the angle between the soft and hard layers using the Slonczewski formula,  {as well as discussing the possible sources of error for this calibration and their magnitude.} Section S5 provides an estimation for the magneto-anisotropic energy barrier between the [110] and [100] magnetization directions, normalized per volume or per atom.  {Section S6 numerically evaluates the possible domain walls pinning by superconducting vortices.} Section S7 discusses the contribution of the shape to the magnetic anisotropy. Finally, section S8 provides details on the theoretical modelling of the observed effects.

\section*{{S1. Magnetic characterizarion of the hard Fe/Co layer}}

\begin{figure}[h]
\begin{center}
\includegraphics[width=\linewidth]{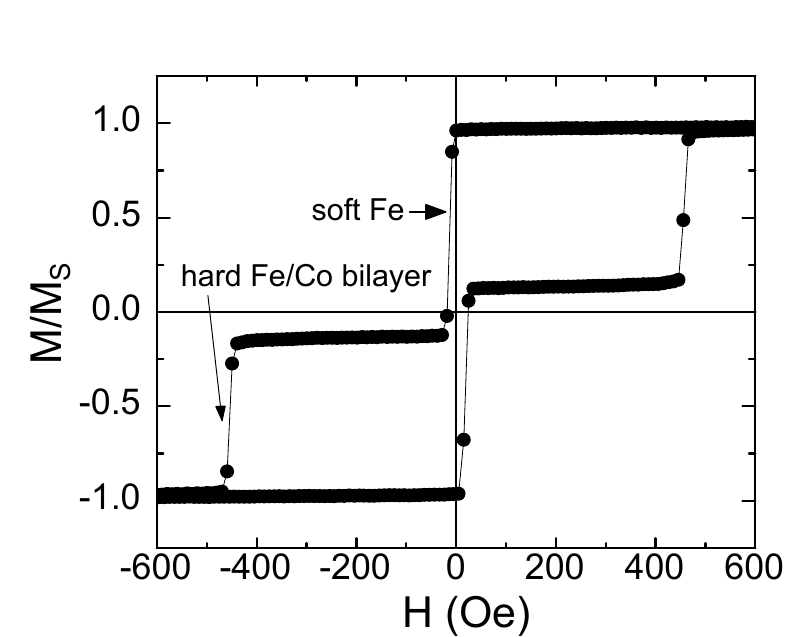}
\caption{ {Magnetic characterization of a Fe(30 nm)/MgO/Fe(10 nm)Co(20 nm) structure, at room temperature, along the [100] direction.}}
\label{FigS1}
\end{center}
\end{figure}

\begin{figure}[h]
\begin{center}
\includegraphics[width=0.9\linewidth]{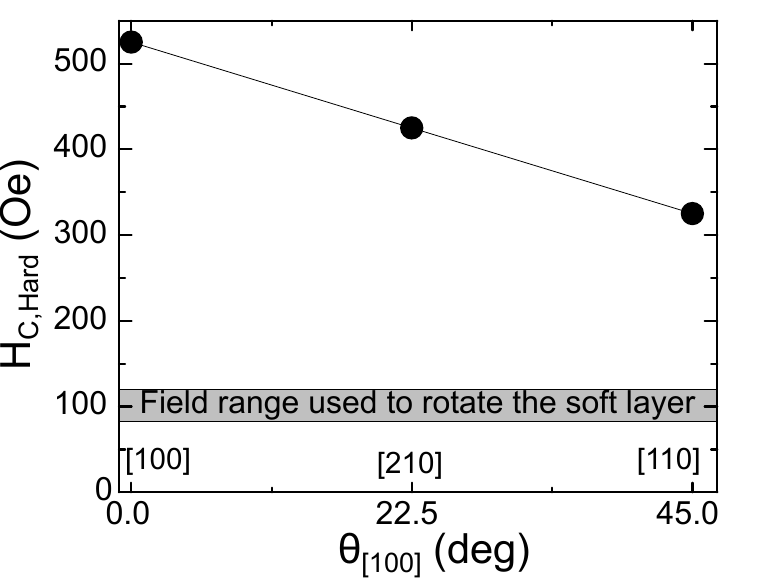}
\caption{ {Coercive field of the hard Fe/Co layer for magnetic field oriented along different crystallographic directions [100], [110] and [210], above the superconducting critical temperature ($T=5$ K). The grey band shows the typical field range used to manipulate the magnetization of the soft Fe(100) layer in the rotation experiments.}}
\label{FigS2}
\end{center}
\end{figure}

  {Figure S\ref{FigS1} shows the magnetic characterization of the hard Fe/Co bilayer, determined from a typical spin-valve M-H loop on a standard Fe/MgO/Fe/Co single crystal MTJ system (continuous layers, unpatterned). The nominal thickness of the layers on this sample, MgO(100)/Fe(30 nm)/MgO(2 nm)/Fe(10 nm)/Co(20 nm), has been chosen to optimize the magnetic properties of the MTJ stack \cite{Tiusan2007}.
The TMR measurements of the coercive fields of the hard (H$_{C,Hard}$) and soft (H$_{C,Soft}$) layers in MTJs under study show that they are well separated from the external field values used to rotate the soft layer}. Figure S\ref{FigS2} shows that the hard layer switching fields obtained from TMRs along [100], [010] and [110] measured in our junctions remain far above the typical range of 70-120 Oe which is used to rotate the soft layer. Moreover, Figure S\ref{FigS3} also shows the typical temperature dependence of H$_{C,Hard}$, demonstrating its independence with temperature from well above to well below $T_C$.

\section*{{S2. Magnetic characterizarion of the soft Fe(001) layer and estimation of the Meissner screening}}

 {The magnetostatic Meissner screening has been discussed mainly in studies with perpendicular magnetization \cite{Bulaevskii2000}. In the case of the experiments with in-plane field rotation which we carry out, such field expulsion could induce some screening of the external magnetic field applied to invert or rotate the magnetization of the soft Fe(001) layer (which is the closest to the superconductor), and with less probability affect the switching of the more distant hard Fe/Co layer.}

\begin{figure}[h]
\begin{center}
\includegraphics[width=0.8\linewidth]{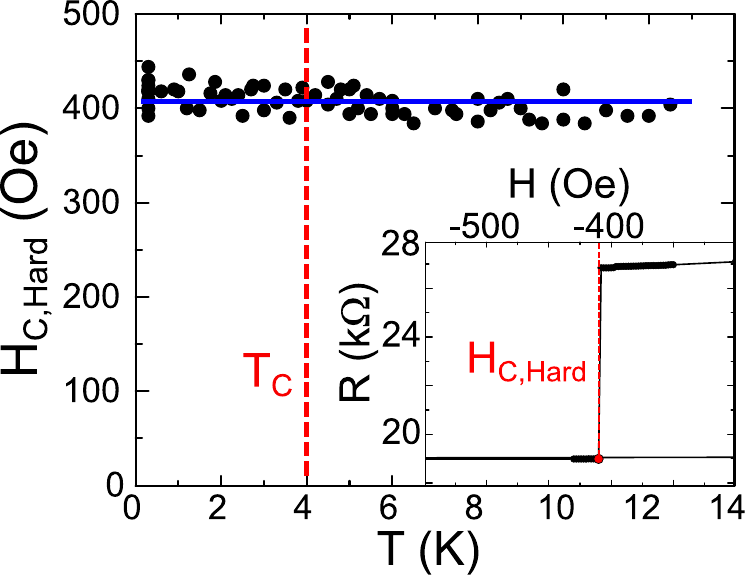}
\caption{Typical temperature dependence of the coercitive field of the hard Fe/Co ferromagnetic layer along the [100] direction. The critical temperature is marked with a dashed red line.  {We relate the excess scatter observed in the hard layer with the extra structural disorder at the Fe/Co interface, providing an enhanced coercive field for the Fe/Co layer. The blue line is a guide for the eye. The inset shows the method for determining the coercive field: it's the field of the first point after the hard layer transition from in each TMR experiment.}}
\label{FigS3}
\end{center}
\end{figure}
\begin{figure}[h]
\begin{center}
\includegraphics[width=0.8\linewidth]{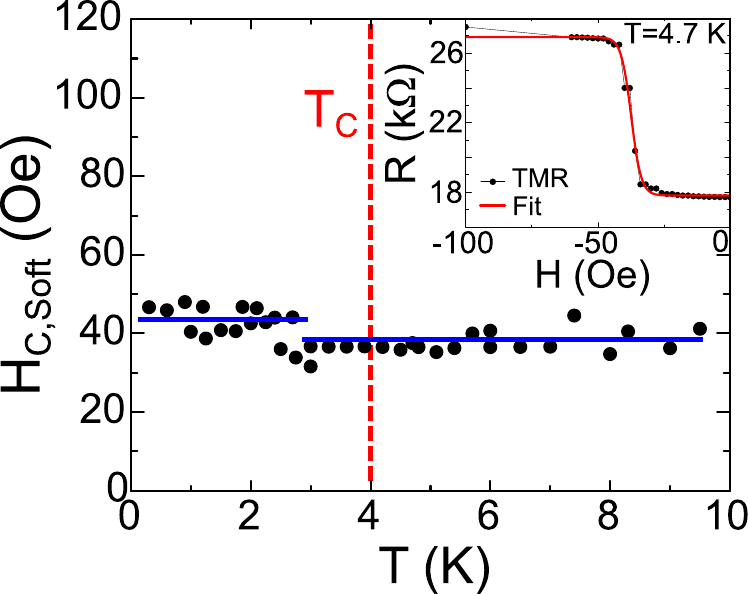}
\caption{Typical temperature dependence of the coercitive field of the soft Fe(001) ferromagnetic layer measured along [100] direction. The critical temperature is marked with a dashed red line.  {Blue lines are guides for the eye. The inset shows the method for determining the coercive field: a logistic fit was performed for the transition, and the coercive field was defined as the mid-height value of the fit.}}
\label{FigS4}
\end{center}
\end{figure}

{Figure S\ref{FigS4} shows the typical variation of the coercive field of the soft Fe(001) ferromagnetic layer with temperature from above to below the critical temperature. We observe some weak increase of the coercive field below 10 Oe, which could be due to spontaneous Meissener screening and/or vortex interaction with domain walls. These changes, however, are an order of magnitude below the typical magnetic fields applied to rotate the Fe(001) layer (70-120 Oe). As we also show in Figure S\ref{FigS3}, the coercive field of the hard FeCo layer (typically above 400-500 Oe) shows practically no variation (within the error bars) within a wide temperature range, from $3T_C$ to $0.1T_C$, discarding the influence of the Meissner screening on the hard layer.}

 {As the superconducting layer is much larger in area than the ferromagnetic one, these experiments point out that the possible existing Meissner screening would introduce about a $10\%$ correction to the actual external field acting on the soft ferromagnet, regardless of the external field direction.}

\section*{{S3. Estimation of the weak antiferromagnetic coupling of the two ferromagnetic layers.}}

\begin{figure}[h]
\begin{center}
\includegraphics[width=0.75\linewidth]{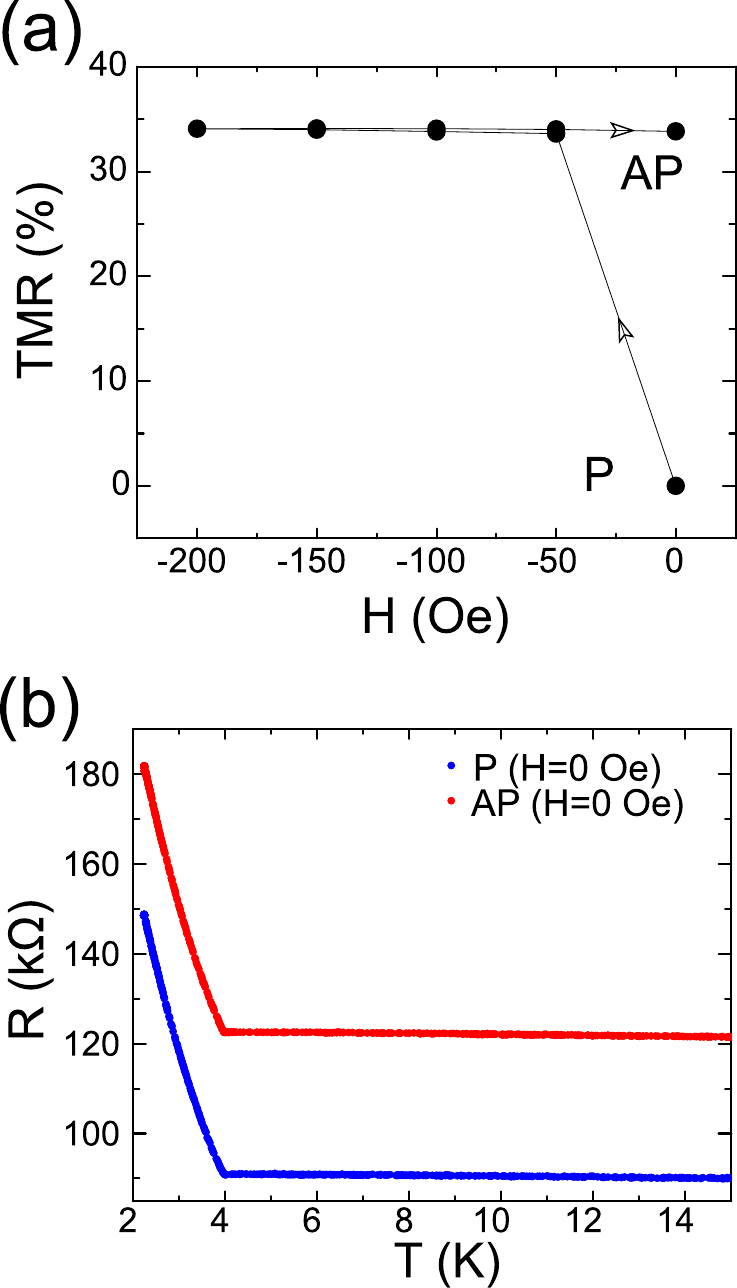}
\caption{ {Two experiments demonstrating the stability of the P and AP states at zero field. (a) TMR to AP state before a critical temperature measurement: the sample was first saturated in the P state with $H=1000$ Oe in the [100] direction, and then a negative field sweeping was performed to -200 Oe and back to 0 Oe in the same direction in order to switch the soft layer into the AP state, where it remained at zero field. (b) Two critical temperature measurements: the sample was saturated in the P state, and then switched to AP state as described in (a) for the AP measurement. After this, the temperature was risen to 15 K and let to slowly cool down to $T\sim2$ K. The increase in resistance below 4 K corresponds to the opening and deepening of the superconducting gap, since the voltage used was only a few microvolts in order to distinguish the superconducting transition from its appearance. Both experiments show no sudden changes in resistance, as would happen if any magnetic transition took place.}}
\label{FigS5}
\end{center}
\end{figure}

 {In order to quantify the unavoidable weak antiferromagnetic magnetostatic coupling between the rotated soft Fe(001) and the practically fixed hard FeCo layer, we show low field TMR measurements where the AP state is achieved and then maintained at zero field (Figure S\ref{FigS5}a). One clearly observes that the AP and P states can be obtained as two different non-volatile states, and therefore the antiferromagnetic coupling is not sufficient to antiferromagnetically couple the two layers at zero field. The stability of the P state against the antiferromagnetic coupling is confirmed by the temperature dependence of the resistance in the P and AP states. The P state shows stable resistance values at least below 15 K (Figure S\ref{FigS5}b). This means that the antiferromagnetic coupling energy is well below 2 mV.}

\section*{S4. Calibration of the angle between the two ferromagnetic layers}

In order to estimate the angle between the two ferromagnets for the TMR measurements and rotations, we used the Slonczewski model \cite{Slonczewski}. By using values of the resistance in the AP, P and PIP states established above $T_\mathrm{C}$, we can calculate the desired angle $\theta$ with the following expression:

\begin{equation}
G^{-1}={G_1}^{-1}+\left[G_2\left(1+p^2\cos{\theta}\right)\right]^{-1}.
\end{equation}

Here, $G$ is the total conductance of the sample, $G_1$ and $G_2$ are the conductances of each of the two tunnel barriers, and $p$ is the spin polarization in the ferromagnets, for which we obtain values between 0.7 and 0.8 depending on the sample  {(the value being robust for each individual one).}

 {In order to ascertain the precission of this calibration method, an analysis of the different errors has been performed. First, an standard error propagation calculation was done to estimate the uncertainty in the resistance values, taking typical values for the current and voltage of 100 nA and 5 mV, respectively, which gives us a typical resistance value of 50 k$\Omega$. The current is applied using a Keithley 220 Current Source, which has an error of $0.3\%$ in the operating range according to the user manual. The voltage is measured using a DMM-522 PCI multimeter card. In the specifications, the voltage precision is said to be 5 $1/2$ digits. With all this, the resistance error obtained is $\Delta$R=75.08 $\Omega$ or a 0.15$\%$ of relative error. Using this value, the error bars in the measurements shown in the main text would be well within the experimental points.}

\begin{figure}[h]
\begin{center}
\includegraphics[width=0.75\linewidth]{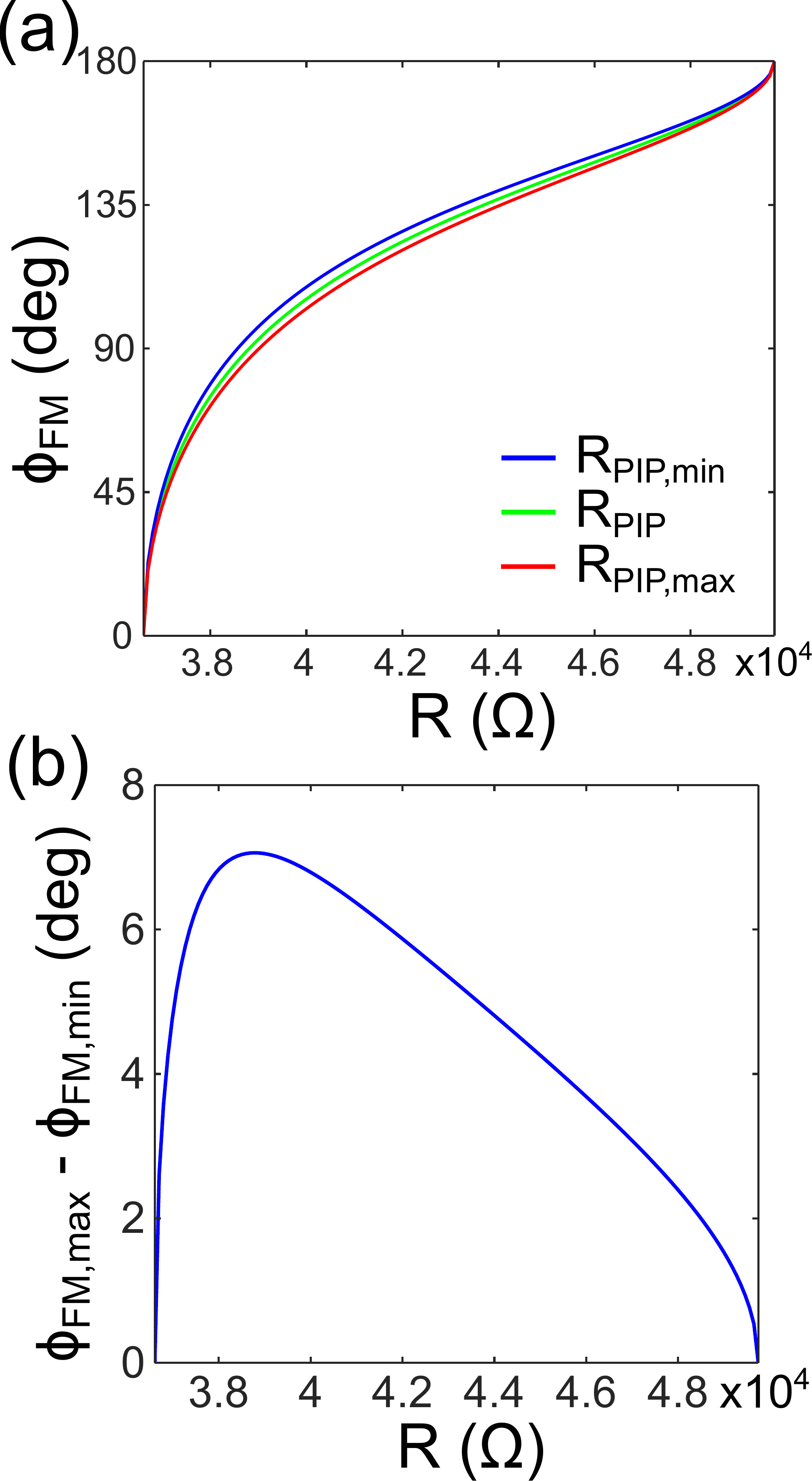}
\caption{ {(a) $\Phi_{FM}$ as a function of resistance for the fittings with maximum, usual, and minimum $R_{PIP}$ used, in the P-AP resistance range. (b) difference of calculated angle vs resistance (in the P-AP resistance range) for the fittings with maximum and minimum $R_{PIP}$ used.}}
\label{FigS6}
\end{center}
\end{figure}

 {For the calculated angle, the error propagation method is not adequate. It gives errors bigger than 360 degrees for some angles, and in general over 30 degrees. This is clearly not what it is observed in reality: the performed fits are quite robust, showing little variance in the estimated angle when changing the input parameters all that is reasonable. Instead, we have used a typical rotation performed on a $30\times30$ $\mu\text{m}^2$ sample. The fitting to the Slonczewski formula needs three input values: the resistance in the P state ($R_P$), the resistance in the AP state ($R_{AP}$) and the resistance in the PIP state ($R_{PIP}$). Using these, a numerical algorithm calculates the spin polarization ($p$), the resistance of the F/F barrier ($R_{FIF}$), and the resistance of the F/S (F/N) barrier ($R_{NIF}$). These give us the total resistance of the sample as a function of the angle $\Phi_{FM}$ between the two ferromagnets or, reciprocally, the angle as a function of resistance. For our estimation, we have varied the value of the $R_{PIP}$ input parameter from the lowest to the highest possible in the PIP state of the rotation, as well as taking an intermediate value which would be used in a normal analysis (the P and AP resistance values are always taken as the minimum and maximum resistance values in the rotation respectively). The calculated parameters for the resistance of each barrier and the polarization may slightly vary from one fitting to another, but the overall fitting remains remarkably stable, as shown in Figure \ref{FigS6}.} 

{As expected, the difference is higher for the PIP state, and minimum in the P and AP state that are ``fixed''. The difference doesn't exceed 7 degrees, and it keeps below 2 degrees near the P and AP states.}

\section*{S5. Saturation magnetization for thin Fe(001) films in [100] and [110] directions}

\begin{figure}[h]
\begin{center}
\includegraphics[width=0.75\linewidth]{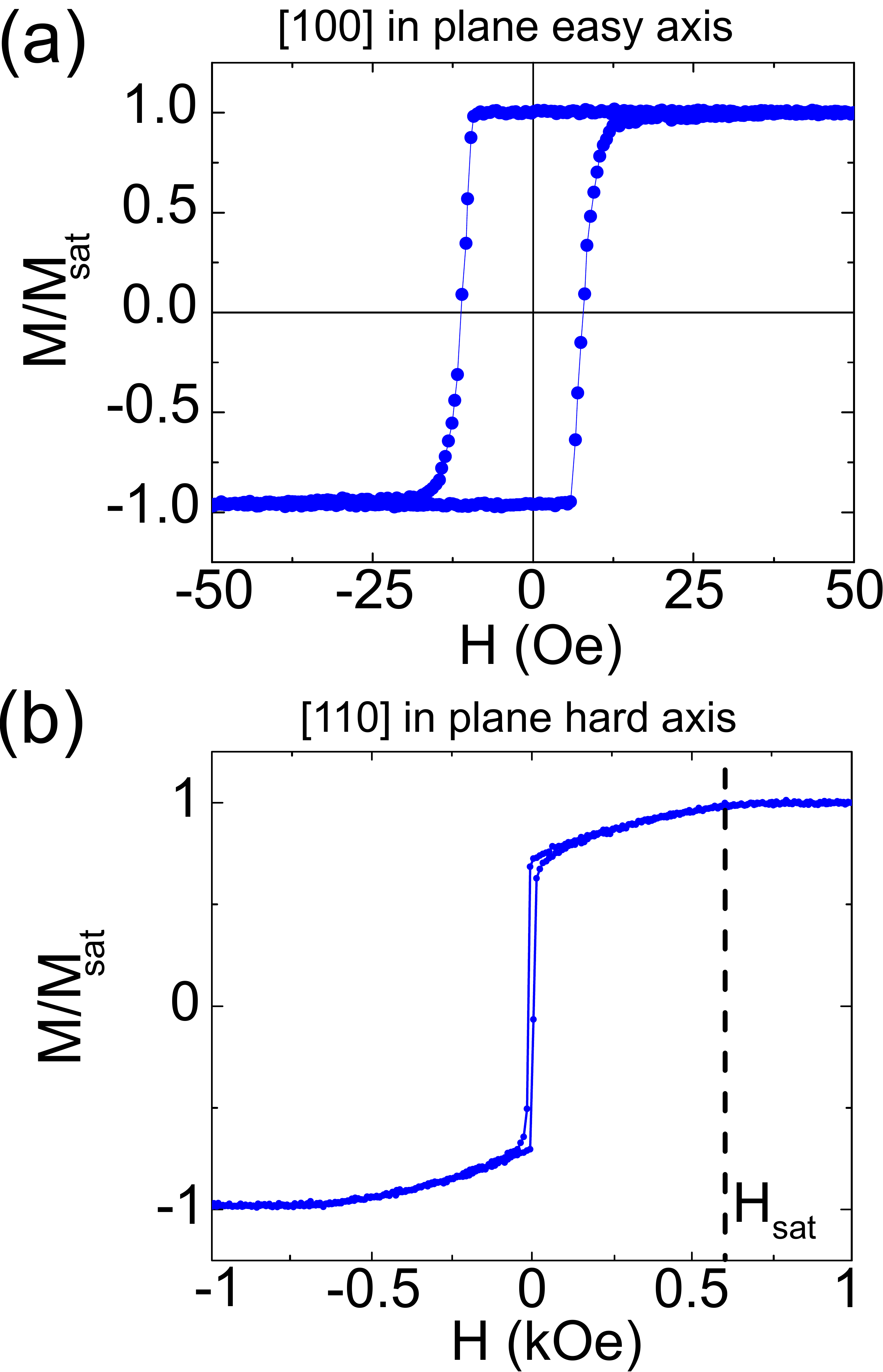}
\caption{M vs H measurements on a 10 nm thick Fe film for the easy [100] (a) and hard [110] (b) crystallographic axis. The saturation field ($H_{sat}$) for the easy axis is around 10 Oe, while for the hard direction it reaches up to 600 Oe. }
\label{FigS7}
\end{center}
\end{figure}

Different M vs H measurements were performed at room temperature on a 10 nm thick Fe films, both for the easy [100] and hard [110] crystallographic axes, in order to estimate the magnetocrystalline anisotropy (MCA) energy. The results are depicted in Fig. S\ref{FigS7}. Using the saturation field for the two directions, the anisotropy energy can be estimated as $K_{Fe}=M_{Fe}H_{Sat}/2=5.1\times10^5\text{ erg $\cdot$ cm}^{-3}$, where $M_{Fe}=1714\text{ emu/cm}^3$ is used. The anisotropy energy per unit cell is therefore MAE$=6.674$ $\mu$eV, or 3.337 $\mu$eV per atom. The obtained energy barrier is similar to the one measured using ferromagnetic resonance \cite{Anisimov1999}. 

\begin{figure}[h]
\begin{center}
\includegraphics[width=0.8\linewidth]{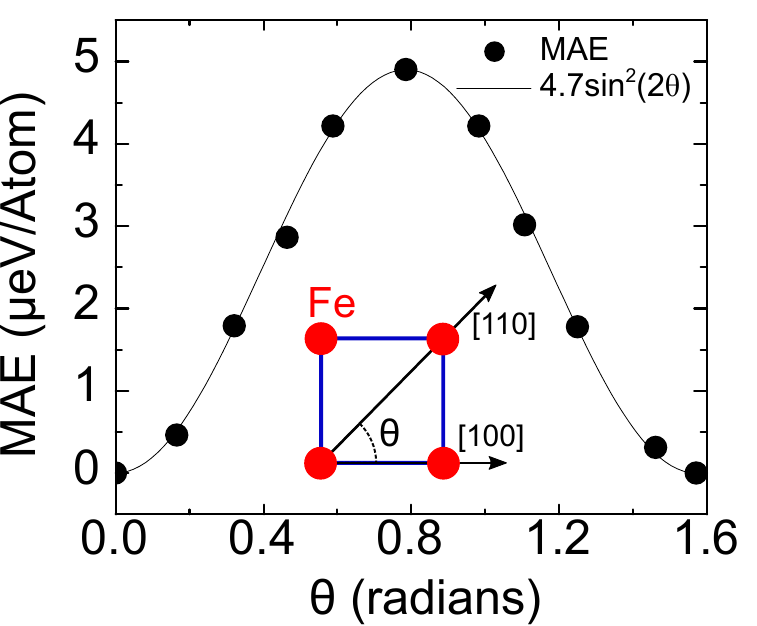}
\caption{Ab-initio calculation of magnetocryscalline anisotropy energy (MAE) as a function of the in-plane orientation angle $\theta$, defined in the inset. Solid line is a phenomenological fit to a $\sin^2(2\theta)$ function.}
\label{FigS8}
\end{center}
\end{figure}

\begin{figure*}[]
\begin{center}
\includegraphics[width=2\columnwidth]{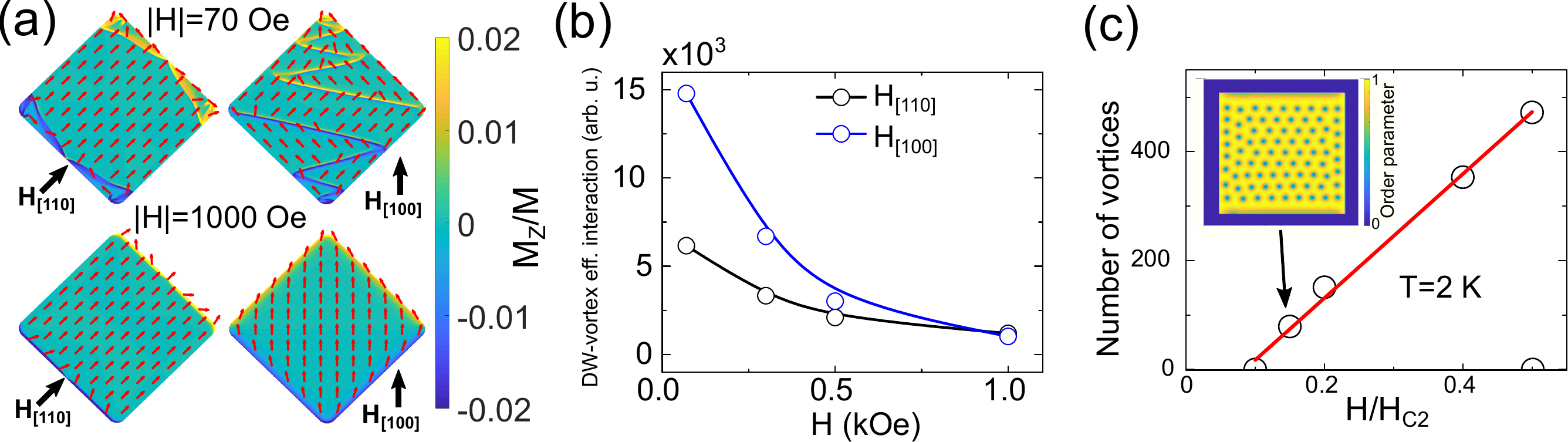}
\caption{{(a) Typical DW formation mapped by MuMax3 simulations for the [110] and [100] applied field directions in the non-saturated (70 Oe) and saturated (1000 Oe) field regimes. The color map represents the out of plane component of the magnetization, while the red arrows indicate the in plane direction. (b) Values of the 2D integral $\mathcal I$ between the local exchange energy (DWs) and the perpendicular component of the stray fields at a distance of 2-3 nm from the ferromagnet, taking into acount the vortex generation function $\mathcal F$. (c) Vortex generation function $\mathcal F$, represented as number of vortices formed in a $5\times5$ $\mu m^2$ square 40 nm thick superconducting Vanadium film as a function of the applied perpendicular field (normalized by the second critical field $H_{c2}$), simulated at $T=2$ K by using the TDGL code described in \cite{Lara2020}. The insert shows a typical image of the vortices at H=0.15$H_{c2}$}}
\label{FigS9}
\end{center}
\end{figure*}

As shown in Fig.S\ref{FigS8}, the experimental MCA energy values have been theoretically confronted with theoretical/numerical calculations of the angular in-plane variation of magnetic anisotropy, using the ab-initio Wien2k FP-LAPW code \cite{Wien2k}. The calculations were based on a supercell model for a V/MgO/Fe/MgO slab similar to the experimental samples. To insure the requested extreme accuracy in MCA energy values ($\mu$eV energy range), a thoroughly well-converged $k$ grid with significantly large number of $k$-points has been involved. Within these circumstances, our theoretical results for the Fe(001) thin films show standard fourfold anisotropy features and reasonable agreement with the experimentally estimated figures with a maximum theoretical MAE of 4.9 $\mu$eV per atom (expected theoretical under-estimation of the magnetocristalline energy within the GGA approach). Note that the superonducting-V induced MCA modulation features cannot be described within the ab-initio FP-LAPW approach, describing the V in its normal metallic state. Therefore, the below $T_\mathrm{C}$ experimentally observed MCA energy modulations have to be clearly related to the proximity effect in the superconducting V/MgO/Fe(001) system and not to any specific MCA feature of Fe(001) in the V/MgO/Fe(001)/MgO complex stacking sequence.

\section*{{S6. Evaluation of the vortex induced pinning of domain walls}}

 {Using MuMax3 \cite{MuMax3}, we have compared numerically the DWs formation along the [100] and [110] magnetization directions. The simulations took place in samples with $3\times3$ $\mu$m$^2$ lateral dimensions (100 nm rounded corners were used as the devices have been fabricated by optical lithography), with $512\times512\times16$ cells, at $T=0$. The rest of the parameters used were $A_{ex}=2.1\times10^{-11}$ J/m for the exchange energy, $M_{\text{sat}}=1.7\times10^6$ A/m for the saturation magnetization, a damping parameter $\alpha=0.02$, and crystalline anisotropy parameters $K_{C1}=4.8\times10^4$ J/m$^3$ and $K_{C3}=-4.32\times10^5$ J/m$^3$. The goal of the simulations was to evaluate the DW formation and their interaction with the superconducting vortices induced by the vertical component of the stray fields at a 2-3 nm from the Fe(001) surface. We observed that, depending on the external field, in the range of 70-1000 Oe both edge-type and inner-type DWs are formed when the field is directed along [100], and mainly edge type DWs are formed with field along [110] (Figure S\ref{FigS9}a).}

 {We have also calculated the interaction $\mathcal I$ between the DW related excess exchange energy $E_{\text{ex}}$ and the vertical component of the stray fields, $B_{\text{eff}}$ (Figure S\ref{FigS9}b):}

\begin{equation}
	 {\mathcal I=\int_{0}^{N_x}\int_{0}^{N_y}\left|B_{\text{eff}}\right|E_{\text{ex}}\mathcal Fdxdy}
\end{equation}

 {Where $N_x$ and $N_y$ are the total number of cells in each dimension of the simulation, and $\mathcal F$ is a filter ``Vortex generation function'' that takes into account the simulated dependence of the number of vortices on the vertically applied field (Figure S\ref{FigS9}c). The vortices were simulated using the Time Dependent Ginzburg Landau code developed in Madrid described in \cite{Lara2020}. The TDGL simulations took place in $5\times5$ $\mu\text{m}^2$ Vanadium samples with $200\times200$ cells, at $T=2$ K, with a coherence length $\xi_0=2.6\times10^{-8}$ based on our experimental estimations for the studied devices, $\kappa=3$ and $T_C=4$ K. A uniform field was applied in the perpendicular direction, its magnitude varying from $0.1H_{C2}$ to $0.6H_{C2}$, and the number of vortices generated in the relaxed state were counted.}

{The second critical field in the vertical direction ($H_{c2} = 3$ kOe) was determined experimentally. The estimated interaction shows that in the weakly saturated regime, when the inner DWs could emerge and the DW-vortex interaction increases, such interaction should pin the magnetization along the [100] direction, corresponding to the MCA already present in the normal state, therefore blocking any magnetization rotation towards the [110] direction, contrary to our experimental observations. The possible reason for the irrelevance of the DW-vortex interaction in our system is that inner DWs are expected to be of Neel-type for the thickness considered \cite{Hubert1999}.}

{Finally we mention that our numerical evaluations show that, if present, the vortex-DW interaction should remain dominant for the magnetization directed along [100] respect [110] and for the magnetic field range 70-1000 Oe also without $K_{C3}$ parameter providing the MCA energy minima along [110]. 
 {In these simulations the Fe(001) layer has been considered to be smooth. In order to further approach simulations to the experiment, we have also verified that the conclusions above are not affected by the introduction of interfacial magnetic disorder due to mismatch defects (every 30 lattice periods) with 25$\%$ excess of Fe moment at the Fe/MgO interface \cite{Jal2015}. More detailed simulations involving also interface roughness could be needed to further approach the real experimental situation.}

\begin{figure*}[]
\begin{center}
\includegraphics[width=2\columnwidth]{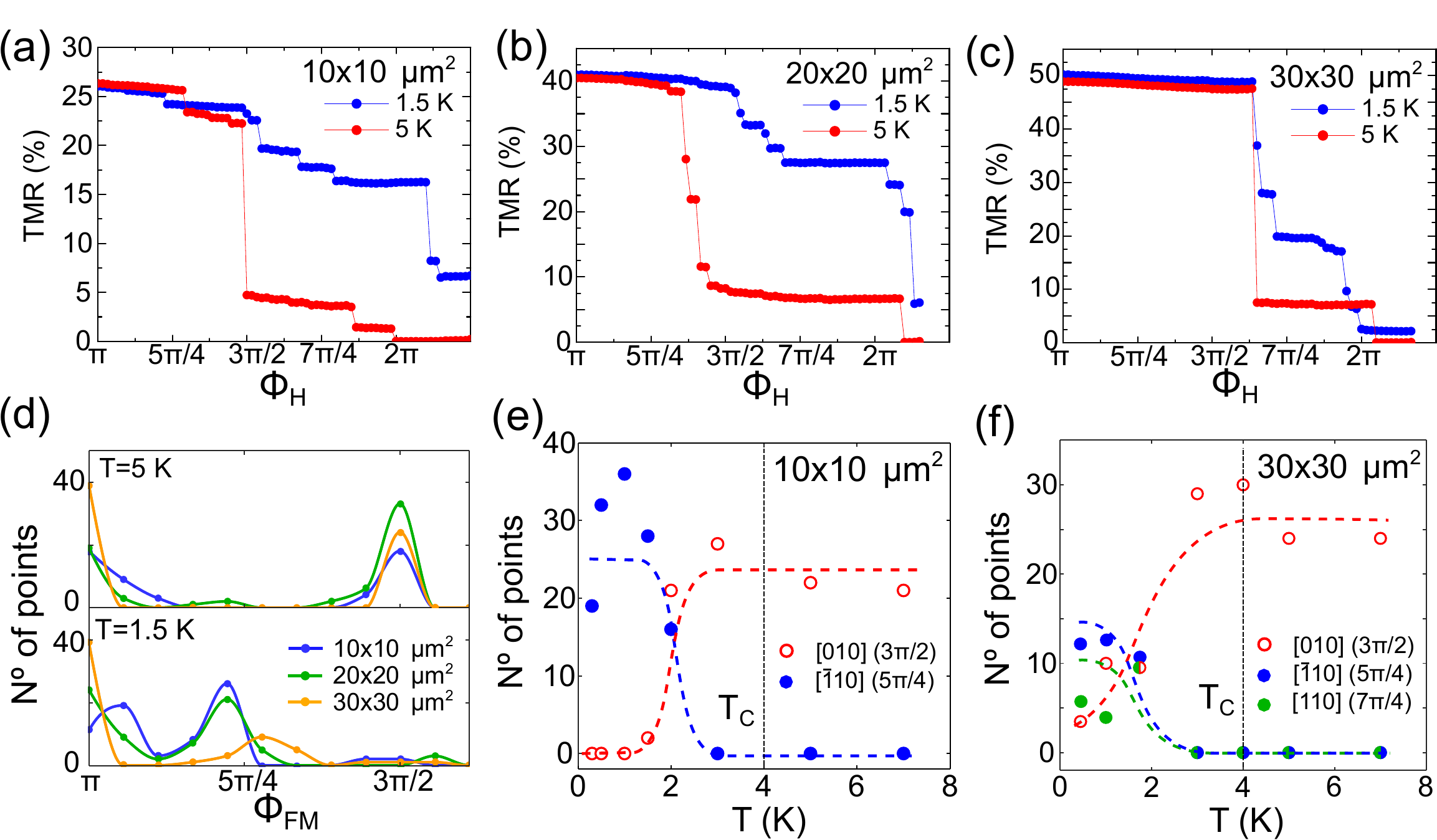}
\caption{In-plane field rotation experiments with $H=70$ Oe below (blue) and above (red) $T_C$ for $10\times10$ (a), $20\times20$ (b) and $30\times30$ (c) $\mu$m$^2$ junctions. (d) Histograms of the calculated angle between the two FM layers $\phi_{FM}$ for these same rotations, above and below $T_\mathrm{C}$, showing the [$\overline1$10] states for low temperatures.  {(e) and (f) show the evolution of the [110], [$\overline1$10] and [010] (PIP) states with temperature for the same $10\times10$ and $30\times30$ $\mu$m$^2$ junctions (qualitatively simular evolution is shown in Figure 2f in the main text for the $20\times20$ $\mu$m$^2$  junction).}}
\label{FigS10}
\end{center}
\end{figure*}

\section*{S7.  {Magnetization alignment along [110] and} irrelevance of the junction area for the  {superconductivity induced} MCA modification}

As we mentioned in the main text, our experiments point that Fe(001) layers are close to a highly saturated state when the magnetization is directed along [100] or equivalent axes. On the other hand, micromagnetic simulations (Figure S\ref{FigS9}a) show that the magnetization alignment is more robust in the [110] direction (or equivalent) rather than in the [100] direction (or equivalent). So, if we indeed reach a highly saturated state in the [100] direction, this should also be the case for the [110] direction. Therefore, the emergent stable tunneling magnetoresistance states we observe experimentally below $T_c$, cannot be explained in terms of the intermediate multi-domain states but rather correspond to the dominant [110] magnetization alignment of the Fe(001) layer.

As shown in Figure S\ref{FigS10}, our experiments shows that the observed effects remain qualitatively unchanged when the junction area is varied about an order of magnitude.

\section*{S8. Modelling}

We describe the V/MgO/Fe structure by the Hamiltonian \cite{Johnsen2019}
\begin{equation}
    \begin{split}
        \label{Hamiltonian}
        H =& -t\sum_{\left<\boldsymbol{i},\boldsymbol{j}\right>,\sigma}c_{\boldsymbol{i},\sigma}^\dagger c_{\boldsymbol{j},\sigma}
        -\sum_{\boldsymbol{i},\sigma} (\mu_{\boldsymbol{i}}-V_{\boldsymbol{i}}) c_{\boldsymbol{i},\sigma}^\dagger c_{\boldsymbol{i},\sigma} \\
        &-\sum_{\boldsymbol{i}} U_{\boldsymbol{i}}n_{\boldsymbol{i},\uparrow}n_{\boldsymbol{i},\downarrow}+\sum_{\boldsymbol{i},\alpha,\beta}c_{\boldsymbol{i},\alpha}^{\dagger}(\boldsymbol{h}_{\boldsymbol{i}}\cdot\boldsymbol{\sigma})_{\alpha,\beta} c_{\boldsymbol{i},\beta}\\
        &-\frac{i}{2}\sum_{\left< \boldsymbol{i},\boldsymbol{j}\right> ,\alpha,\beta} \lambda_{\boldsymbol{i}} c_{\boldsymbol{i},\alpha}^{\dagger} \hat{n} \cdot (\boldsymbol{\sigma}\times\boldsymbol{d}_{\boldsymbol{i},\boldsymbol{j}})_{\alpha,\beta} c_{\boldsymbol{j},\beta}\\
    \end{split}
\end{equation}
defined on a cubic lattice. The first term describes nearest-neighbor hopping. The second term includes the the chemical potential and the potential barrier at the insulating MgO layers. The remaining terms describes superconducting attractive on-site interaction, ferromagnetic exchange interaction, and Rashba spin-orbit interaction, respectively. These are only nonzero in their respective regions. In the above, $t$ is the hopping integral, $\mu_{\boldsymbol{i}}$ is the chemical potential, $V_{\boldsymbol{i}}$ is the potential barrier that is nonzero only for the MgO layer, $U>0$ is the attractive on-site interaction giving rise to superconductivity, $\lambda_{\boldsymbol{i}}$ is the local spin-orbit coupling magnitude, $\hat{n}$ is a unit vector normal to the interface, $\boldsymbol{\sigma}$ is the vector of Pauli matrices, $\boldsymbol{d}_{\boldsymbol{i},\boldsymbol{j}}$ is a vector from site $\boldsymbol{i}$ to site $\boldsymbol{j}$, and $\boldsymbol{h_i}$ is the local magnetic exchange field.
The number operator used above is defined as $n_{\boldsymbol{i},\sigma}\equiv c_{\boldsymbol{i},\sigma}^\dagger c_{\boldsymbol{i},\sigma}$, and  $c_{\boldsymbol{i},\sigma}^\dagger$ and $c_{\boldsymbol{i},\sigma}$ are the second-quantization electron creation and annihilation operators at site $\boldsymbol{i}$ with spin $\sigma$.
The superconducting term in the Hamiltonian is treated by a mean-field approach, where we assume $c_{\boldsymbol{i},\uparrow} c_{\boldsymbol{i},\downarrow} = \left< c_{\boldsymbol{i},\uparrow} c_{\boldsymbol{i},\downarrow} \right> +\delta$ and neglect terms of second order in the fluctuations $\delta$.

We consider a system of size $N_x \times N_y \times N_z$ setting the interface normals parallel to the $x$ axis and assuming periodic boundary conditions in the $y$ and $z$ directions. To simplify notation in the following, we define $i\equiv i_x$, $j\equiv j_x$, $\boldsymbol{i}_{||}=(i_x , i_y )$ and $\boldsymbol{k}\equiv (k_y, k_z)$. We apply the Fourier transform
\begin{equation}
    \label{FT}
    c_{\boldsymbol{i},\sigma}=\frac{1}{\sqrt{N_y N_z}}\sum_{\boldsymbol{k}} c_{i,\boldsymbol{k},\sigma} e^{\mathrm{i}(\boldsymbol{k}\cdot\boldsymbol{i}_{||})}
\end{equation}
to the above Hamiltonian and use that
\begin{equation}
    \label{rel}
    \frac{1}{N_y N_z}\sum_{\boldsymbol{i}_{||}} e^{\mathrm{i}(\boldsymbol{k}-\boldsymbol{k}')\cdot\boldsymbol{i}_{||}}=\delta_{\boldsymbol{k} , \boldsymbol{k}'}.
\end{equation}
We choose a new basis
\begin{equation}
    B_{i , \boldsymbol{k}}^{\dagger}=[c_{i , \boldsymbol{k} ,\uparrow}^{\dagger} \hspace{2mm} c_{i , \boldsymbol{k} ,\downarrow}^{\dagger} \hspace{2mm} c_{i , -\boldsymbol{k} ,\uparrow} \hspace{2mm} c_{i , -\boldsymbol{k} ,\downarrow}]
\end{equation}
spanning Nambu$\times$spin space, and rewrite the Hamiltonian as
\begin{equation}
\label{H2}
    H=H_0+\frac{1}{2}\sum_{i , j , \boldsymbol{k}} B_{i , \boldsymbol{k}}^\dagger H_{i , j , \boldsymbol{k}}B_{i , \boldsymbol{k}}.
\end{equation}
Above, the Hamiltonian matrix is given by
\begin{equation}
\label{Hamiltonian2}
    \begin{split}
        H_{i , j , \boldsymbol{k}}&= \epsilon_{i ,j ,\boldsymbol{k}} \hat{\tau}_3 \hat{\sigma}_0 +\delta_{i ,j}\Big[\mathrm{i}\Delta_{i}\hat{\tau}^+ \hat{\sigma}_y  -\mathrm{i}\Delta_{i}^* \hat{\tau}^- \hat{\sigma}_y \\
        &+h_{i}^x \hat{\tau}_3 \hat{\sigma}_x +h_{i}^y \hat{\tau}_0 \hat{\sigma}_y +h_{i}^z \hat{\tau}_3 \hat{\sigma}_z\\
        & -\lambda_{i}\sin(k_y )\hat{\tau}_0 \hat{\sigma}_z +\lambda_{i}\sin(k_z )\hat{\tau}_3 \hat{\sigma}_y \Big],\\
    \end{split}
\end{equation}
where $\Delta_i$ is the superconducting gap which we solve for self-consistently, $\hat{\tau}_i \hat{\sigma}_j \equiv \hat{\tau}_i \otimes\hat{\sigma}_j$ is the Kronecker product of the Pauli matrices spanning Nambu and spin space, $\hat{\tau}^\pm \equiv(\hat{\tau}_1 \pm i\hat{\tau}_2 )/2$, and
\begin{equation}
\begin{split}
    \epsilon_{i , j ,  \boldsymbol{k}} \equiv& -2t\left[\cos(k_y)+\cos(k_z)\right]\delta_{i , j}\\
    &-t(\delta_{i , j +1}+\delta_{i , j -1})\\
    &-(\mu_{i}-V_i )\delta_{i , j}.
\end{split}
\end{equation}
The constant term in Eq.~\eqref{H2} is given by
\begin{equation}
\begin{split}
    H_0=&- \sum_{i , \boldsymbol{k}}\left\{2t\left[\cos(k_y)+\cos(k_z)\right]+\mu_{i}-V_i \right\} \\
    &+N_y N_z \sum_{i}\frac{|\Delta_{i}|^2}{U_{i}}.
\end{split}
\label{H0}
\end{equation}
We absorb the sum over lattice sites in Eq.~\eqref{H2} into the matrix product by defining a new basis
\begin{equation}
    W_{\boldsymbol{k}}^\dagger = [B_{1,\boldsymbol{k}}^\dagger,...,B_{i ,\boldsymbol{k}}^\dagger ,...,B_{N_x ,\boldsymbol{k}}^\dagger ].
\end{equation}
Eq.~\eqref{H2} can then be rewritten as 
\begin{equation}
    H=H_0 + \frac{1}{2}\sum_{\boldsymbol{k}}W_{\boldsymbol{k}}^\dagger H_{\boldsymbol{k}} W_{\boldsymbol{k}},
\end{equation}
where
\begin{equation}
\label{Hkykz}
    H_{\boldsymbol{k}}=
    \begin{bmatrix}
        H_{1,1,\boldsymbol{k}} & \cdots & H_{1,N_x ,\boldsymbol{k}}\\
        \vdots & \ddots & \vdots \\
        H_{N_x ,1,\boldsymbol{k}} & \cdots & H_{N_x ,N_x ,\boldsymbol{k}}\\
    \end{bmatrix}
\end{equation}
is Hermitian and can be diagonalized numerically. We obtain eigenvalues $E_{n,\boldsymbol{k}}$ and eigenvectors $\Phi_{n,\boldsymbol{k}}$ given by 
\begin{equation}
\begin{split}
    \Phi_{n,\boldsymbol{k}}^{\dagger}&=[ \phi_{1,n,\boldsymbol{k}}^{\dagger} \hspace{3mm} \cdots \hspace{3mm}\phi_{N_x ,n,\boldsymbol{k}}^{\dagger}],\\
    \phi_{i ,n,\boldsymbol{k}}^{\dagger}&=[u_{i ,n,\boldsymbol{k}}^{*}\hspace{1mm} v_{i ,n,\boldsymbol{k}}^{*}\hspace{1mm} w_{i ,n,\boldsymbol{k}}^{*}\hspace{1mm} x_{i ,n,\boldsymbol{k}}^{*}].
\end{split}
\end{equation} 
The diagonalized Hamiltonian can be written on the form
\begin{equation}
    H=H_0+\frac{1}{2}\sum_{n, \boldsymbol{k}}E_{n, \boldsymbol{k}}\gamma_{n, \boldsymbol{k}}^\dagger \gamma_{n, \boldsymbol{k}},
\end{equation}
where the new quasi-particle operators are related to the old operators by
\begin{equation}
\label{c}
\begin{split}
     c_{i_ ,\boldsymbol{k} ,\uparrow}&=\sum_n u_{i_ ,n, \boldsymbol{k}}\gamma_{n,\boldsymbol{k}},\\
    c_{i ,\boldsymbol{k} ,\downarrow}&=\sum_n v_{i ,n, \boldsymbol{k}}\gamma_{n,\boldsymbol{k}},\\
    c_{i ,-\boldsymbol{k} ,\uparrow}^{\dagger}&=\sum_n w_{i ,n, \boldsymbol{k}}\gamma_{n,\boldsymbol{k}},\\
    c_{i ,-\boldsymbol{k} ,\downarrow}^{\dagger}&=\sum_n x_{i ,n, \boldsymbol{k}}\gamma_{n,\boldsymbol{k}}.
\end{split}
\end{equation}

The superconducting gap is given by $\Delta_{\boldsymbol{i}}\equiv U_{\boldsymbol{i}}\left<c_{\boldsymbol{i},\uparrow}c_{\boldsymbol{i},\downarrow}\right>$. We apply the Fourier transform in Eq.~\eqref{FT} and use Eq.~\eqref{c} in order to rewrite the expression in terms of the new quasi-particle operators. Also using that $\langle \gamma_{n,\boldsymbol{k}}^\dagger \gamma_{m,\boldsymbol{k}}\rangle =f\big(E_{n,\boldsymbol{k}}/2\big)\delta_{n,m}$, we obtain the expression
\begin{equation}
    \label{deltaIt}
        \Delta_{i}=
        -\frac{U_{i}}{N_y N_z}\sum_{n,\boldsymbol{k}}v_{i ,n,\boldsymbol{k}}w_{i ,n,\boldsymbol{k}}^* \left[1-f\left(E_{n,\boldsymbol{k}}/2\right)\right]
\end{equation}
for the gap, that we use in computing the eigenenergies iteratively.
Above, $f\big(E_{n,\boldsymbol{k}}/2\big)$ is the Fermi-Dirac distribution. 

Using the obtained eigenenergies, we compute the free energy,
\begin{equation}
    \label{F}
    F=H_0 -\frac{1}{\beta}\sum_{n,\boldsymbol{k}}\ln(1+e^{-\beta E_{n, \boldsymbol{k}}/2}),
\end{equation}
where $\beta=(k_B T)^{-1}$. The preferred magnetization directions are described by the local minima of the free energy. In the main body of the paper, we use this to explain the possible magnetization directions of the soft ferromagnet when rotating an IP external magnetic field over a $2\pi$ angle starting at a parallel alignment with the hard ferromagnet. 

Other relevant quantities to consider in modelling the experimental system is the superconducting coherence length and the superconducting critical temperature. 
In the ballistic limit, the coherence length is given by $\xi=\hbar v_F /\pi\Delta_0$, where $v_F=\frac{1}{\hbar}\frac{dE_{\boldsymbol{k}}}{dk}\big|_{k=k_F}$ is the Fermi velocity related to the normal-state eigenenergy $E_{\boldsymbol{k}}=-2t[\cos(k_x )+\cos(k_y )+\cos(k_z )]-\mu$, and $\Delta_0$ is the zero-temperature superconducting gap \cite{Bardeen1957Theory}. 
The critical temperature is found by a binomial search, where we decide if a temperature is above or below $T_c$ by determining whether $\Delta_{N_x^{\text{S}}/2}$ increases towards a superconducting solution or decreases towards a normal state solution from the initial guess under iterative recalculations of $\Delta_{i}$. We choose an initial guess with a magnitude very close to zero and with a lattice site dependence similar to that of the gap just below $T_c$. 

In the main plot showing the free energy under IP rotations of the magnetization, we have chosen parameters $t=1$, $\mu_\text{S} = \mu_\text{{SOC}}=\mu_{\text{F}}=0.9$, $V=2.1$, $U=1.35$, $\lambda=0.4$, $h=0.8$, $N_{x}^{\text{S}}=30$, $N_x^{\text{SOC}}=3$, $N_x^{\text{F}}=8$, and $N_y = N_z = 60$. 
 {All length scales are scaled by the lattice constant $a$, all energy scales are scaled by the hopping parameter $t$, and the magnitude of the spin-orbit coupling $\lambda$ is scaled by $ta$.
In order to make the system computationally manageable, the lattice size is scaled down compared to the experimental system, however the results should give qualitatively similar results as long as the ratios between the coherence length and the layer thicknesses are reasonable compared to the experimental system. For this set of parameters, the superconducting coherence length is approximately $0.6N_x^{\text{S}}$.
Since the coherence length is inversely proportional to the superconducting gap, $U$ has been chosen to be large in order to allow for a coherence length smaller than the thickness of the superconducting layer. 
Although this results in a large superconducting gap, the modelling will qualitatively fit the experimental results as long as the other parameters are adjusted accordingly. We therefore choose the local magnetic exchange field so that $h\gg\Delta$, as in the experiment. 
For this parameter set, $h\approx20\Delta$.
The order of magnitude of $\lambda$ is $1$~eV\AA, given that $t\sim1$~eV and $a\sim4$~\AA. This is realistic considering Rashba parameters measured in several materials \cite{manchon_nm_15}.
The Rashba spin-orbit field at the interfaces of V/MgO/Fe is caused by a structural inversion asymmetry across the MgO layer, and breaks the inversion symmetry at the MgO interfaces \cite{Martinez2020}. This causes generation of triplet-superconductivity even for weakly spin-polarized ferromagnets with a small spin-orbit field \cite{vezin_prb_20}. We are therefore not dependent upon a strong magnetic exchange field and a strong spin-orbit field for realizing the observed effects.
For the AF coupling contribution to the free energy, we set $f_{\text{AF}}=0.01$ in order to fit the anisotropy of the experimental system just above $T_\mathrm{C}$.}


\begin{thebibliography}{99}

\bibitem{Ginsburg1956} V.L. Ginzburg, Ferromagnetic superconductors, Zh. Eksp. \href{http://www.jetp.ac.ru/cgi-bin/dn/e_004_02_0153.pdf}{Teor. Fiz. \textbf{31}, 202 (1956); Sov. Phys. JETP \textbf{4}, 153 (1957).}

\bibitem{Matthias1958} B. T. Matthias, H. Suhl and E. Corenzwit, Spin exchange in superconductors, \href{https://journals.aps.org/prl/abstract/10.1103/PhysRevLett.1.92}{Phys. Rev. Lett. \textbf{1}, 152 (1958).}

\bibitem{Buzdin1984}  A. I. Buzdin, L. N. Bulaevskii, M. L. Kulich and S. V. Panyukov, Magnetic superconductors, \href{https://iopscience.iop.org/article/10.1070/PU1984v027n12ABEH004085/meta}{Sov Phys. Uspekhi, \textbf{27}, 927 (1984).}

\bibitem{Bader2002} J. Y. Gu, C.-Y. You, J. S. Jiang, J. Pearson, Y. B. Bazaliy and S. D. Bader, Magnetization-orientation dependence of the superconducting transition temperature in the ferromagnet-superconductor-ferromagnet system: CuNi/Nb/CuNi, \href{https://journals.aps.org/prl/abstract/10.1103/PhysRevLett.89.267001}{Phys. Rev. Lett. \textbf{89}, 267001 (2002).}

\bibitem{Birge2006} I. C. Moraru, W. P. Pratt, Jr. and N. O. Birge, Observation of standard spin-switch effects in ferromagnet/superconductor/ferromagnet trilayers with a strong ferromagnet, \href{https://journals.aps.org/prb/abstract/10.1103/PhysRevB.74.220507}{Phys. Rev. B \textbf{74}, 220507(R) (2006).}

\bibitem{Tagirov1999} L. R. Tagirov, Low-field superconducting spin switch based on a superconductor/ferromagnet multilayer, \href{https://journals.aps.org/prl/abstract/10.1103/PhysRevLett.83.2058}{Phys. Rev. Lett. \textbf{83}, 2058 (1999).}

\bibitem{Buzdin1999} A. I. Buzdin, A. V. Vedyayev and N. V. Ryzhanova, Spin-orientation dependent superconductivity in F/S/F structures, \href{https://iopscience.iop.org/article/10.1209/epl/i1999-00539-0}{Europhys. Lett., \textbf{48} (6), 686-691 (1999).}

\bibitem{Baladie2001} I. Baladi\'e, A. Buzdin, N. Ryzhanova and A. Vedyayev, Interplay of superconductivity and magnetism in superconductor/ferromagnet structures,\href{https://journals.aps.org/prb/abstract/10.1103/PhysRevB.63.054518}{Phys. Rev. B \textbf{63}, 054518 (2001).}

\bibitem{Leksin2011}  {P. V. Leksin, N. N. Garif'yanov, I. A. Garifullin, J. Schumann, V. Kataev, O. G. Schmidt and B. B\"uchner, Manifestation of New Interference Effects in a Superconductor-Ferromagnet Spin Valve, \href{https://journals.aps.org/prl/abstract/10.1103/PhysRevLett.106.067005}{Phys. Rev. Lett. \textbf{106}, 067005 (2011).}}
 
\bibitem{Wang2014} X. L. Wang, A. Di Bernardo, N. Banerjee, A. Wells, F. S. Bergeret, M. G. Blamire and J. W. A. Robinson, Giant triplet proximity effect in superconducting pseudo spin valves with engineered anisotropy, \href{https://journals.aps.org/prb/abstract/10.1103/PhysRevB.89.140508}{Phys. Rev. B \textbf{89}, 140508(R) (2014).}

\bibitem{Singh2015} A. Singh, S. Voltan, K. Lahabi and J. Aarts, Colossal Proximity Effect in a Superconducting Triplet Spin Valve Based on the Half-Metallic Ferromagnet CrO$_2$, \href{https://journals.aps.org/prx/abstract/10.1103/PhysRevX.5.021019}{Phys. Rev. X \textbf{5}, 021019 (2015).}

\bibitem{Leksin2012} P.V. Leksin, N. N. Garif’yanov, I. A. Garifullin, Ya.V. Fominov, J. Schumann, Y. Krupskaya, V. Kataev, O. G. Schmidt and B. B\"uchner, Evidence for triplet superconductivity in a superconductor-ferromagnet spin valve, \href{https://journals.aps.org/prl/abstract/10.1103/PhysRevLett.109.057005}{Phys. Rev. Lett., \textbf{109}, 057005 (2012).}

\bibitem{Keizer2006} R. S. Keizer, S. T. B. Goennenwein, T. M. Klapwijk, G. Miao, G. Xiao and  A. A. Gupta, A spin triplet supercurrent through the half-metallic ferromagnet CrO$_2$, \href{https://www.nature.com/articles/nature04499}{Nature \textbf{439}, 825 (2006).}

\bibitem{Khaire2010} T. S. Khaire, M. A. Khasawneh, W. P. Pratt, Jr. and N. O. Birge, Observation of spin-triplet superconductivity in Co-based Josephson junctions, \href{https://journals.aps.org/prl/abstract/10.1103/PhysRevLett.104.137002}{Phys. Rev. Lett. \textbf{104}, 137002 (2010).}

\bibitem{Anwar2010} M. S. Anwar, F. Czeschka, M. Hesselberth, M. Porcu and J. Aarts, Long-range supercurrents through half-metallic ferromagnetic  CrO$_2$, \href{https://journals.aps.org/prb/abstract/10.1103/PhysRevB.82.100501}{Phys. Rev. B \textbf{82}, 100501(R) (2010).}

\bibitem{Robinson2010} J. W. A. Robinson, J. D. S. Witt and M. G. Blamire, Controlled injection of spin-triplet supercurrents into a strong ferromagnet, \href{https://science.sciencemag.org/content/329/5987/59.abstract}{Science \textbf{329}, 59 (2010).}

\bibitem{Visani2012} C. Visani, Z. Sefrioui, J. Tornos, C. Leon, J. Briatico, M. Bibes, A. Barth\'el\'emy, J. Santamar\'ia and J. E. Villegas, Equal-spin Andreev reflection and long-range coherent transport in high-temperature superconductor/halfmetallic ferromagnet junctions, \href{https://www.nature.com/articles/nphys2318}{Nature Phys., \textbf{8}, 540 (2012).}

\bibitem{Banerjee2014} N. Banerjee, J. W. A. Robinson and M. G. Blamire, Reversible control of spin-polarized supercurrents in ferromagnetic Josephson junctions, \href{https://www.nature.com/articles/ncomms5771}{Nature Comm. \textbf{5}, 4771 (2014).}

\bibitem{Krasnov2014} A. Iovan, T. Golod and V. M. Krasnov, Controllable generation of a spin-triplet supercurrent in a Josephson spin valve, \href{https://journals.aps.org/prb/abstract/10.1103/PhysRevB.90.134514}{Phys. Rev. B \textbf{90}, 134514 (2014).}

\bibitem{Linder2015} J. Linder and J. W. A. Robinson, Superconducting spintronics, \href{https://www.nature.com/articles/nphys3242?proof=true}{Nature Phys. \textbf{11}, 307 (2015).}

\bibitem{Eschrig2015} M. Eschrig, Spin-polarized supercurrents for spintronics: a review of current progress, \href{https://iopscience.iop.org/article/10.1088/0034-4885/78/10/104501/pdf}{Rep. Prog. Phys. \textbf{78}, 104501 (2015).}

\bibitem{Flokstra2016} M. G. Flokstra, N. Satchell, J. Kim, G. Burnell, P. J. Curran, S. J. Bending, J. F. K. Cooper, C. J. Kinane, S. Langridge, A. Isidori, N. Pugach, M. Eschrig, H. Luetkens, A. Suter, T. Prokscha and S. L. Lee, Remotely induced magnetism in a normal metal using a superconducting spin-valve, \href{https://www.nature.com/articles/nphys3486}{Nature Phys., \textbf{12}, 97 ( 2016).}

\bibitem{Buzdin1988}  {A.I. Buzdin and L. N. Bulaevskii, Ferromagnetic film on the surface of a superconductor: Possible onset of inhomogeneous magnetic ordering, \href{http://jetp.ac.ru/cgi-bin/e/index/e/67/3/p576?a=list}{Zh. Eksp. Teor. Fiz. \textbf{94}, 256-261 (1988), Sov. Phys. JETP \textbf{67}, 576-578 (1988).}}

\bibitem{Bulaevskii2000}  {L. N. Bulaevskii and E. M. Chudnovsky, Ferromagnetic film on a superconducting substrate, \href{https://journals.aps.org/prb/abstract/10.1103/PhysRevB.63.012502}{Phys. Rev. B \textbf{63}, 012502 (2000).}}

\bibitem{Dubonos2002}  {S. V. Dubonos, A. K. Geim, K. S. Novoselov and I. V. Grigorieva, Spontaneous magnetization changes and nonlocal effects in mesoscopic ferromagnet-superconductor structures, \href{https://journals.aps.org/prb/abstract/10.1103/PhysRevB.65.220513}{Phys. Rev. B \textbf{65}, 220513R (2002).}}

\bibitem{Fritzsche2009}  {J. Fritzsche, R. B. G. Kramer and V. V. Moshchalkov, Visualization of the vortex-mediated pinning of ferromagnetic domains in superconductor-ferromagnet hybrids, \href{https://journals.aps.org/prb/abstract/10.1103/PhysRevB.79.132501}{Phys. Rev. B \textbf{79}, 132501 (2009).}}

\bibitem{Curran2015}  {P. J. Curran et al., Irreversible magnetization switching at the onset of superconductivity in a superconductor ferromagnet hybrid, \href{https://aip.scitation.org/doi/10.1063/1.4938467}{Appl. Phys. Lett. \textbf{107}, 262602 (2015).}}

\bibitem{Waintal2002}
 {X. Weintal and P. W. Brouwer, Magnetic exchange interaction induced by a Josephson current, \href{https://journals.aps.org/prb/abstract/10.1103/PhysRevB.65.054407}{Phys. Rev. B \textbf{65}, 054407 (2002).}}

\bibitem{Tserkovnyak2002}
 {Y. Tserkovnyak and A. Brataas, Current and spin torque in double tunnel barrier ferromagnet-superconductorferromagnet systems, \href{https://journals.aps.org/prb/abstract/10.1103/PhysRevB.65.094517}{Phys. Rev. B \textbf{65}, 094517 (2002).}}


\bibitem{Bell2008}
 {C. Bell, S. Milikisyants, M. Huber and J. Aarts, Spin dynamics in a superconductor-ferromagnet proximity system, \href{https://journals.aps.org/prl/pdf/10.1103/PhysRevLett.100.047002}{Phys. Rev. Lett. \textbf{100}, 047002 (2008).}}

\bibitem{Braude2008}
 {B. Braude and Ya. M. Blanter, Triplet Josephson effect with magnetic feedback in a superconductor-ferromagnet heterostructure, \href{https://journals.aps.org/prl/abstract/10.1103/PhysRevLett.100.207001}{Phys. Rev. Lett. \textbf{100}, 207001 (2008).}}

\bibitem{Zhao2008}
 {E. Zhao and J. A. Sauls, Theory of nonequilibrium spin transport and spin-transfer torque in superconducting-ferromagnetic nanostructures, \href{https://journals.aps.org/prb/abstract/10.1103/PhysRevB.78.174511}{Phys. Rev. B \textbf{78}, 174511 (2008).}}

\bibitem{Konschelle2009}
 {F. Konschelle and A. Buzdin, Magnetic moment manipulation by a Josephson current, \href{https://journals.aps.org/prl/abstract/10.1103/PhysRevLett.102.017001}{Phys. Rev. Lett. \textbf{102}, 017001 (2019).}}

\bibitem{Linder2011}
 {J. Linder and T. Yokoyama, Supercurrent-induced magnetization dynamics in a Josephson junction with two misaligned ferromagnetic layers, \href{https://journals.aps.org/prb/abstract/10.1103/PhysRevB.83.012501}{Phys. Rev. B \textbf{83}, 012501 (2011).}}

\bibitem{Linder2012}
 {J. Linder, A. Brataas, Z. Shomali and M. Zareyan, Spin-transfer and exchange torques in ferromagnetic superconductors, \href{https://journals.aps.org/prl/abstract/10.1103/PhysRevLett.109.237206}{Phys. Rev. Lett. \textbf{109}, 237206 (2012).}}

\bibitem{Pugach2012}
 {N. G. Pugach and A. I. Buzdin, Magnetic moment manipulation by triplet Josephson current, \href{https://aip.scitation.org/doi/full/10.1063/1.4769900}{Appl. Phys. Lett. \textbf{101}, 242602 (2012).}}

\bibitem{Bergeret2013} F. S. Bergeret and I. V. Tokatly, Singlet-triplet conversion and the long-range proximity effect in superconductor-ferromagnet structures with generic spin dependent fields, \href{https://journals.aps.org/prl/abstract/10.1103/PhysRevLett.110.117003}{Phys. Rev. Lett., \textbf{110}, 117003 (2013).}

\bibitem{Banerjee2018} N. Banerjee, J. A. Ouassou, Y. Zhu, N. A. Stelmashenko, J. Linder and M. G. Blamire, Controlling the superconducting transition by spin-orbit coupling, \href{https://journals.aps.org/prb/abstract/10.1103/PhysRevB.97.184521}{Phys. Rev. B \textbf{97}, 184521 (2018).}

\bibitem{Jeon2018} {K.-R. Jeon, C. Ciccarelli, A. J. Ferguson, H. Kurebayashi, L. F. Cohen, X. Montiel, M. Eschrig, J. W. A. Robinson and M. G. Blamire, Enhanced spin pumping into superconductors provides evidence for superconducting pure spin currents, \href{https://doi.org/10.1038/s41563-018-0058-9}{Nat. Mater. \textbf{17}, 499 (2018).}}

\bibitem{Jeon2019} {K.-R. Jeon, C. Ciccarelli, A. J. Ferguson, H. Kurebayashi, L. F. Cohen, X. Montiel, M. Eschrig, S. Komori, J. W. A. Robinson and M. G. Blamire, Exchange-field enhancement of superconducting spin pumping, \href{https://doi.org/10.1103/PhysRevB.99.024507}{Phys. Rev. B \textbf{99}, 024507 (2019).}}

\bibitem{Satchell2018}  {N. Satchell and N. O. Birge, Supercurrent in ferromagnetic Josephson junctions with heavy metal interlayers, \href{https://doi.org/10.1103/PhysRevB.97.214509}{Phys. Rev. B \textbf{97}, 214509 (2018).} }

\bibitem{Satchell2019}  {N. Satchell, R. Loloee and N. O. Birge, Supercurrent in ferromagnetic Josephson junctions with heavy-metal interlayers. II. Canted magnetization, \href{https://doi.org/10.1103/PhysRevB.99.174519}{Phys. Rev. B \textbf{99}, 174519 (2019).}}

\bibitem{Jacobsen2015} S. H. Jacobsen, J. A. Ouassou and J. Linder, Critical temperature and tunneling spectroscopy of superconductor-ferromagnet hybrids with intrinsic Rashba-Dresselhaus spin-orbit coupling, \href{https://journals.aps.org/prb/abstract/10.1103/PhysRevB.97.184521}{Phys. Rev. B \textbf{92}, 024510 (2015).}

\bibitem{Martinez2020} I. Mart\'inez, P. H\"ogl, C. Gonz\'alez-Ruano, J. P. Cascales, C. Tiusan, Y. Lu, M. Hehn, A. Matos-Abiague, J. Fabian, I. Zutic and F. G. Aliev, Interfacial spin-orbit coupling: a platform for superconducting spintronics,  \href{https://journals.aps.org/prapplied/abstract/10.1103/PhysRevApplied.13.014030}{Phys. Rev. Appl. \textbf{13}, 014030 (2020).}

\bibitem{Johnsen2019} L. G. Johnsen, J. Linder and N. Banerjee, Magnetization reorientation due to superconducting transition in heavy metal heterostructures, \href{https://journals.aps.org/prb/abstract/10.1103/PhysRevB.99.134516}{Phys. Rev. B \textbf{99}, 134516 (2019).}

\bibitem{Yang2011} H. X. Yang, M. Chshiev and B. Dieny, First-principles investigation of the very large perpendicular magnetic anisotropy at Fe/MgO and Co/MgO interfaces. \href{https://journals.aps.org/prb/abstract/10.1103/PhysRevB.84.054401}{Phys. Rev. B \textbf{84}, 054401 (2011).}

\bibitem{Martinez2018} I. Mart\'inez, C. Tiusan, M. Hehn, M. Chshiev and F. G. Aliev, Symmetry broken spin reorientation transition in epitaxial MgO/Fe/MgO layers with competing anisotropies, \href{https://www.nature.com/articles/s41598-018-27720-7}{Sci. Rep. \textbf{8}, 9463 (2018).}

\bibitem{Tiusan2007} C. Tiusan, M. Hehn, F. Montaigne, F. Greullet, S. Andrieu and A. Schuhl, Spin tunneling phenomena in single crystal magnetic tunnel junction systems, \href{https://iopscience.iop.org/article/10.1088/0953-8984/19/16/165201/meta}{J. Phys.: Condens. Matter \textbf{19}, 165201 (2007).}

\bibitem{SuppMat} See Supplemental Material at XXX for more details.

\bibitem{Parkin2004} {S. S. P. Parkin, C. Kaiser, A. Panchula, P. M. Rice, B. Hughes, M. Samant, and S.-H. Yang, Giant tunnelling magnetoresistance at room temperature with MgO (100) tunnel barriers, \href{https://www.nature.com/articles/nmat1256}{Nat. Mater. \textbf{3}, 862 (2004).}}

\bibitem{Yuasa2004} {S. Yuasa, T. Nagahama, A. Fukushima, Y. Suzuki and K. Ando, Giant room temperature magneto-resistance in single-crystal Fe/MgO/Fe magnetic tunnel junctions, \href{https://www.nature.com/articles/nmat1257}{Nat. Mater. \href{3}, 868 (2004).}}

\bibitem{MuMax3} {A. Vansteenkiste, J. Leliaert, M. Dvornik, M. Helsen, F. Garcia-Sanchez and B. Van Waeyenberge, The design and verification of MuMax3, \href{https://doi.org/10.1063/1.4899186}{AIP Advances \textbf{4}, 107133 (2014).}}

\bibitem{Lara2020} {A. Lara, C. Gonz\'alez-Ruano and F.G.Aliev, Time-dependent Ginzburg-Landau simulations of superconducting vortices in three dimensions, \href{https://doi.org/10.1063/10.0000861}{Low Temperature Physics \textbf{46}, 316 (2020).}}

\bibitem{Herranz2010} D. Herranz, F. Bonell, A. Gomez-Ibarlucea, S. Andrieu, F. Montaigne, R. Villar, C. Tiusan, and F.G.Aliev, Strongly suppressed 1/f noise and enhanced magnetoresistance in epitaxial Fe–V/MgO/Fe magnetic tunnel junctions, \href{https://aip.scitation.org/doi/10.1063/1.3430064}{Applied Physics Letters \textbf{96}, 202501 (2010).}

\bibitem{Vezin2020} {T. Vezin, C. Shen, J. E. Han and I. \v{Z}uti\'c, Enhanced spin-triplet pairing in magnetic junctions with s-wave superconductors, \href{https://doi.org/10.1103/PhysRevB.101.014515}{Phys. Rev. B \textbf{101}, 014515 (2020).}

\bibitem{Slonczewski} J. C. Slonczewski, Conductance and exchange coupling of two ferromagnets separated by a tunneling barrier. \href{https://journals.aps.org/prb/abstract/10.1103/PhysRevB.39.6995}{Phys. Rev. B \textbf{39}, 6995 (1989).}

\bibitem{Anisimov1999} A. N. Anisimov, M. Farle, P. Poulopoulos, W. Platow, K. Baberschke, P. Isberg, R. Wäppling, A. M. N. Niklasson and O. Eriksson, Orbital Magnetism and Magnetic Anisotropy Probed with Ferromagnetic Resonance, \href{https://journals.aps.org/prl/abstract/10.1103/PhysRevLett.82.2390}{Phys. Rev. Lett. \textbf{82}, 2390 (1999).}

\bibitem{Wien2k} P. Blaha, K.Schwarz, F. Tran, R. Laskowski, G.K.H. Madsen and L.D. Marks, An APW+lo program for calculating the properties of solids, \href{https://aip.scitation.org/doi/10.1063/1.5143061}{J. Chem. Phys. \textbf{152}, 074101 (2020).}

\bibitem{Lara2020} {A. Lara, C. Gonz\'alez-Ruano and F.G.Aliev, Time-dependent Ginzburg-Landau simulations of superconducting vortices in three dimensions, \href{https://doi.org/10.1063/10.0000861}{Low Temperature Physics \textbf{46}, 316 (2020).}}

\bibitem{Hubert1999} A. Hubert and R. Schäfer, Magnetic domains. The analysis of magnetic microstructures (Springer, Berlin, 1999).

\bibitem{Jal2015} E. Jal et al; Interface Fe magnetic moment enhancement in MgO/Fe/MgO trilayers, \href{https://doi.org/10.1063/1.4929990}{Appl. Phys. Lett. \textbf{107}, 092404 (2015).}

\bibitem{Bardeen1957Theory} J. Bardeen, L. N. Cooper and J. R. Schrieffer, Theory of Superconductivity, \href{https://journals.aps.org/pr/abstract/10.1103/PhysRev.108.1175}{Phys. Rev. \textbf{108}, 1175 (1957).}

\bibitem{manchon_nm_15} A. Manchon, H. C. Koo, J. Nitta, S. M. Frolov, and R. A. Duine, New perspectives for Rashba spin–orbit coupling, \href{https://www.nature.com/articles/nmat4360}{Nat. Mater. \textbf{14}, 871 (2015).}

\bibitem{vezin_prb_20} T. Vezin, C. Shen, J. E. Han, and I. \v{Z}uti\'{c}, Enhanced spin-triplet pairing in magnetic junctions with $s$-wave superconductors, \href{https://journals.aps.org/prb/abstract/10.1103/PhysRevB.101.014515}{Phys. Rev. B, \textbf{101}, 014515 (2020).}
}
\end{thebibliography}
\end{document}